\documentclass[aps,pre,twocolumn,superscriptaddress,groupedaddress,showpacs]{revtex4-1}
\usepackage{amsmath,amssymb,graphicx,bbold}
\usepackage{graphicx}  
\usepackage{dcolumn}   
\usepackage{bm}        
\usepackage{color}
\usepackage{subfigure}
\usepackage{cleveref}
\usepackage{amsmath}    
\usepackage{graphics}
\usepackage{verbatim}   
\usepackage{color}      
\usepackage{subfigure}  
\usepackage{hyperref}   
\usepackage{lineno} 
\usepackage{textcomp}

\begin{document}

\title{Correlations in Electrically Coupled Chaotic Lasers}

\author{E. J. Rosero$^1$,  W. A. S. Barbosa$^1$, J. F. Martinez Avila$^{1,2}$, 
A. Z. Khoury$^{1,3}$, J. R. Rios Leite$^1$}

\affiliation{
1- Departamento de F\'{\i}sica,~Universidade Federal de Pernambuco,\\
50670-901 Cidade Universit\'aria, Recife, PE, Brazil;\\ 
2- Departamento de F\'{\i}sica - Universidade Federal de Sergipe,\\
Av. Marechal Rondon, S/N Jardim Rosa Elze - 49100-000 - S\~ao Crist\'ov\~ao - SE - Brazil;\\
3- Instituto de F\'{\i}sica - Universidade Federal Fluminense,\\
Av. Gal. Milton Tavares de Souza S/N, 24210-346 - Niteroi - RJ - Brazil}

\date{\today}
\begin{abstract}

We show how two electrically coupled semiconductor lasers having optical feedback can 
present simultaneous anti-phase correlated fast power fluctuations, 
and strong in-phase synchronized spikes of chaotic power drops.
This quite counter intuitive phenomenon is demonstrated experimentally and 
confirmed by numerical solutions of a deterministic dynamical system of rate equations.
The occurrence of negative and positive cross correlation between parts of a complex system 
according to time scales, as 
proved in our simple arrangement, is relevant for the understanding and characterization of 
collective properties  in complex networks.

\end{abstract}


\maketitle

\section{introduction}

For a long time, light emitting diodes are known to show photon emission correlations 
depending on their electrical pumping coupling. When parallel connected and pumped by 
a very regular current source their output have negative cross-correlation \cite{Edwards92}. 
For pairs of diode lasers this quantum optics effect extends into classical anti-phase 
fluctuations in the power 
emitted by the two laser \cite{ElectronCoupleMIT-APL}. 
To explain these effects physically one needs to remember that the pump electrical carriers 
flow and recombine either way between the parallel connected units. 
Quantum correlations in power fluctuation among pairs of lasers with common pump source have also 
been studied many years ago \cite{Zelaquett1999} and one can also get intuitive explanation for their behavior.
The realm of classical nonlinear dynamics not always  
give such simple intuitive results. 
This will be experimentally shown here for coupled pairs of chaotic diode lasers.
The power fluctuations in the lasers present the coexistence of anti-phase 
fluctuations at a fast time scale simultaneously  with in-phase, fully chaotic synchronized 
power drops in a time scale two orders of magnitude slower.
The occurrence of anti-correlations in subsystems of complex systems that have collective synchronized 
states is an intriguing effect pertaining to different domains in nature. 
One finds it described in economics \cite{economics2014} where the data from anti phase correlated 
pairs of stocks are proposed to extract best conditions for investors to make gains. 
Analogous to what we show for lasers, the  anti-phase 
oscillations of stock values may occur as the market changes and even through the events of crashes, when 
both stocks have simultaneous huge drop.  

In this work we show the dynamics of one pair of lasers chaotic by optical feedback and coupled electrically. 
The results show how competitive coupling for pump energy 
among two chaotic subsystems can lead to synchronized pulse spikes in the whole system while anti phase
oscillations remain present among the parts. 
We can optically and electrically probe the variables and this allows 
detailed experimental inspection of the dynamics and its comparison with theoretical model. 
Data series acquired from each subsystem unit can be statistically matched to numerically calculated data,
extracted from  an autonomous  deterministic 
dynamical system with time delay. The physical interpretation of the equations is available and 
numerical solutions give excellent agreement  with the experiment.  
Such mechanisms can have relevant impact on understanding large laser network dynamics.

From the practical point of view, diode lasers are the most used in optical engineering. 
The nonlinear behavior of a single Edge Emitting diode laser with external cavity optical feedback has many dynamical 
forms as its pump current is changed \cite{threedecadeslff}. Among these are the  so called Low Frequency 
Fluctuation (LFF) in power. The laser acquires a chaotic regime of fast power fluctuations along with strong 
power drops with irregular large time intervals. 
Most of this dynamics can be predicted by a deterministic semiclassical model of rate 
equations with delay \cite{langkobayashi,sano} 
and its experimental study is still attracting broad interest like deterministic coherence resonance \cite{sciamannaEJP2013} and optical rogue events created by feedback with conjugated fields \cite{Mercier2015}. A comprehensive review of laser diode chaos can be found in \cite{shoreNat2015}.  
Modeled as excitable systems \cite{giudiciandronov,eguia99,avila2008,selmi2014} 
they also have been applied to simulate 
complex networks \cite{RMPmirasso,1500laserscoupled}. 

To have a single diode laser presenting chaotic LFF \cite{threedecadeslff,langkobayashi,tartwijklenstraQSC1995}, 
an external mirror placed a few meters apart and
aligned as an external cavity,
feeds back part of the field with a time delay, $\tau$, in the range of tens of nanoseconds.
Then, apart from the optical field period close to $10^{-14}s\,$,
three time scales can be identified in the intensity instabilities \cite{sano,tartwijklenstraQSC1995}: 
First there are ultrafast field fluctuations in the $10$ picosecond range. 
These field amplitude and phase fluctuations 
result from the {\it quasi-mode locking} process among the external cavity modes that creates  ultra-short pulses. 
Next, the laser output power may show fluctuating modulations 
in the intermediate feedback time scale on the order of $10$ nanoseconds, again due to the external reflecting feedback cavity. 
Finally the irregular LFF power drops occur with an average time interval on the 
microseconds to millisecond  range, another two or more orders of magnitude slower. 
These instabilities are reproduced theoretically  with a deterministic set of equations in a dissipative 
nonlinear system \cite{sano}. All these effects are within classical fluctuations scales. 
Light and pump currents quantum fluctuations in the experiments are not addressed and consistently the equations 
are fully deterministic without quantum fluctuation terms. 

Optically coupled pairs of the above described lasers have been studied as 
they present cross correlated dynamics including chaotic synchronization \cite{Ahlers1999,timescalesWu2006},
which has fundamental and applied interest \cite{RMPmirasso,Diodedatatrensmission2005}. 
Novel dynamical behavior appears and are reported here when, instead of optically coupled, 
the pair of lasers have optical feedback but is electronically coupled in parallel.
Examined with broadband time resolution, the dynamics of each laser 
has instabilities in the three above referred time scales. 
The main new result of our work is the demonstration that 
the two-laser system do not show the same type of correlations in the different time scales. 
Synchronism with in phase correlation 
at the slow scale is observed along with anti phase power fluctuations at the intermediate faster scale, 
while no correlation appears in the ultrafast time scale. 
We also show how simple laser rate equations, including 
electric current conservation, match the experiments and opens the possibility for 
new numerical studies in laser networks. 

\section{experimental setup}

\begin{figure}[!hbtp]
 \centering
\includegraphics[width=7.0cm,height=4.5cm]{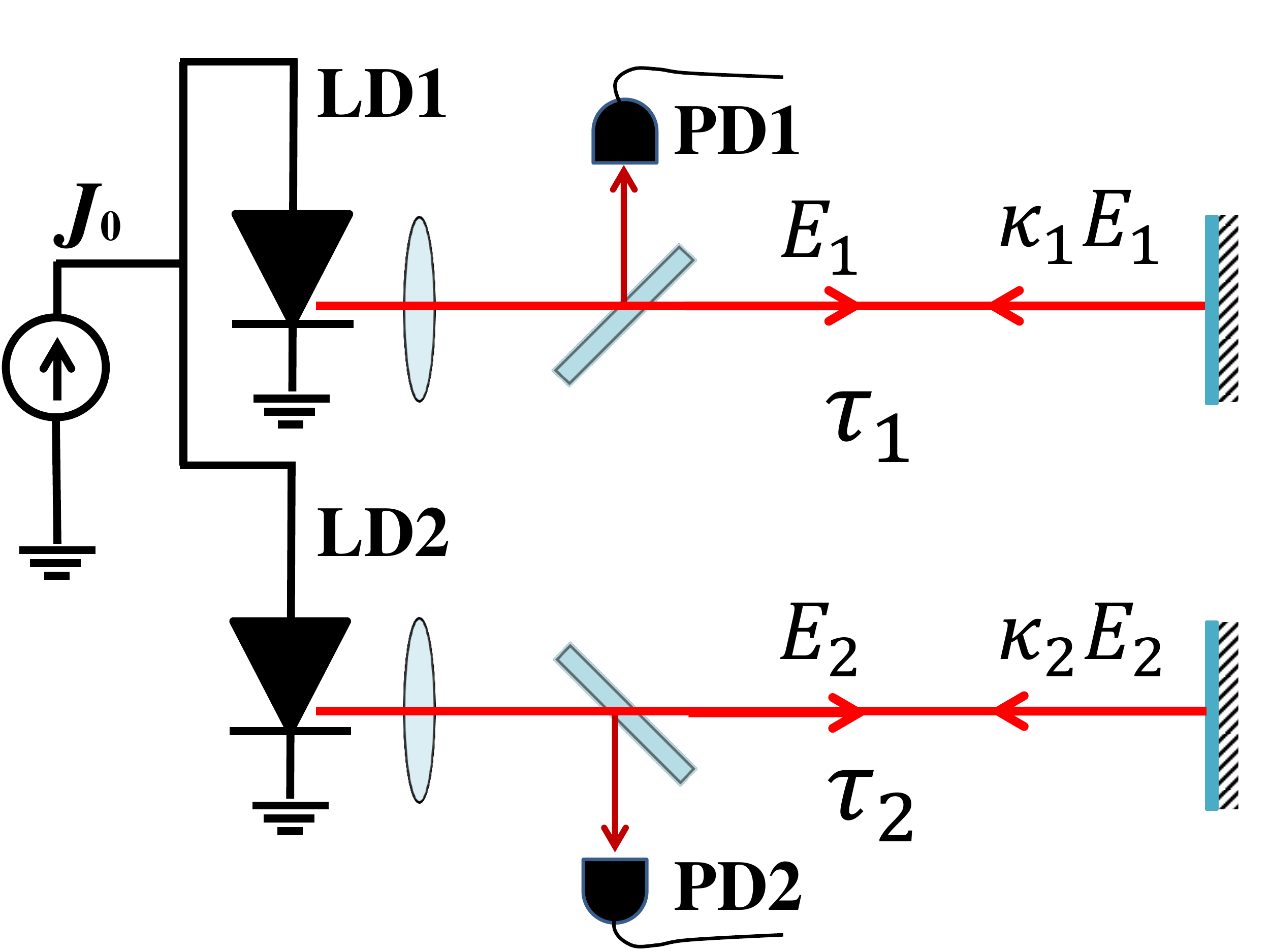}
\caption{Setup for the experiments on the power correlations and chaos
synchronization of two electrically coupled lasers. }
  \label{fig:Fig1}
 \end{figure}

Let us first describe the experimental setup.
Pairs of semiconductor lasers differing by less than $2 \%$ in their 
threshold current and optical 
frequency were coupled electrically in parallel configuration and  
pumped by a high impedance current source, as indicated in Fig. \ref{fig:Fig1}.
%
Various types of commercial GaAlAs single transverse
mode Fabry-Perot laser were used:  one pair of Hitachi-HL8334MG, emitting at $830$ nm with threshold
current of $33$ mA, a pair of ThorLabs L850P010 with $10.5$mA threshold current and $850$ nm
wavelength and another pair of L780P010 with threshold of $8.5$ mA near $780$ nm wavelength.  
All present the same phenomena. 
Solitary longitudinal mode separation was typically $150GHz$ and 
no control was used to keep the lasers mono mode. The electric coupling between the lasers was
made by short coaxial cable connections. 
The total current was stabilized to $\pm0.001$ mA and each laser temperature 
to $0.01$ K.  
Optical feedback was implemented by  
reflecting dielectric mirrors located at a distance of $6.00$ m from each laser, 
after beam collimation by aspheric anti-reflection coated 
lenses.  Manipulating the optical alignment, up to  $6 \%$ current threshold reduction   
could be achieved in both lasers.
The feedback delay times were $\tau_j =40$ ns 
with a $\pm 5$ ps precision mismatch 
that did not affect the results. 
Output coupling beam splitters lead $4 \%$ of light onto 3 GHz
bandwidth photodiodes.
Setting the  total pump current near twice the single laser free running threshold,
makes each laser present 
LFF. Output power data series were acquired by a $1$ GHz bandwidth 
digital oscilloscope. The time series were computer treated to calculate  
experimental histograms and correlation functions. In the section \ref{electrical} 
more is given on the experimental electronic details and measurements made in the system.

\section{Dynamical Equations with Pump Coupling}

The theoretical frame to explain the experiments has the simple form of a rate equations dynamical system
with time delay.
Our model is a pair of coupled equations corresponding to mono mode 
laser equations, having delayed optical field feedback, as introduced by Lang and Kobayashi \cite{langkobayashi,sano}, 
supplemented by Kirchhoff\textquotesingle s laws, with the constraint imposing constant total electronic pump current, $J_0$. 
\begin{eqnarray}
\frac{dE_j}{dt} &=& \frac{\left(1+i \alpha_j \right)}{2} 
\left[G_j(N_j)-\frac{1}{\tau_{pj}}\right] E_j(t) 
 + \kappa_j E_j(t- \tau_j) \;,
 \label{eq:LKeq1}
\nonumber\\
\\
\frac{dN_j}{dt} &=& J_j (t) - \frac{N_j(t)}{\tau_{sj}}- G_j(N_j) \left|E_j(t)\right|^{2} \;,
\label{eq:LKeq2}
\end{eqnarray}
where the gain for each laser $(j=1,2)$ is given by
\begin{eqnarray}
G_j(N_j)=\frac{G_{0j}[N_j(t)-N_{0j}]}{1+ \epsilon_j \left|E_j(t)\right|^{2}}\;.
\label{eq:LKeq3}
\end{eqnarray}
In these equations the dynamical variables are the complex optical electric field, $E_j(t)$, and the 
carrier population inversion, $N_j(t)$.  
The parameters are: the diode laser 
linewidth enhancement factor $\alpha_j$, the photon lifetime in the laser chip cavity $\tau_{pj}$, the optical 
feedback strength $\kappa_j$ and the feedback delay times $\tau_j$. 
The small signal gain is $G_{0j}$, the transparency population inversion is $N_{0j}$ and the gain
saturation coefficient $\epsilon_j$. 
With the parallel coupled circuit configuration the total pump current is split among the lasers as  
$J_0=J_1(t)+J_2(t)$.
The variations of each current are assumed to depend 
linearly on the carrier population difference. Therefore
\begin{equation} 
J_1(t)= J_0/2 -\eta[N_1(t)-N_2(t)]
\end{equation}
and $J_2(t)= J_0-J_{1}(t)$. The value of the coupling coefficient $\eta$ is 
determined from our experimental data taken with a single laser having feedback.  
The threshold pump currents are given by $J_{thj}= (N_{0j}+(\tau_{pj}\,\ G_{0j})^{-1}){\ /}\tau_{sj}$.
The two lasers are similar so that, in most calculations, we assumed the parameters summarized in Table \ref{parameters}.   
They have been attributed values according to early studies 
\cite{tartwijklenstraQSC1995,langkobayashi,sano} and had fine adjustments by inspection of our experimental data. 
The pump current was taken as $J_0 = 2.03J_{th1}$.
\begin{table}[hbtp]
\begin{tabular}{|c|c||c|c|}
\hline
$\alpha_j$ & $3.0$ & $ \ G_{0j}$ & $1.2\times10^{4}\ s^{-1}$ \\ \hline
$\ N_{01}$ & $1.0\times10^{8}$ & $\ N_{02}$ & $0.99\times10^{8}$  \\ \hline
1/$\tau_{pj} $& $513\times10^{9}\ s^{-1}$ & 1/$\tau_{sj}$ & $0.5\times10^{9}\ s^{-1}$ \\ \hline
$\kappa_{1}=\kappa_{2}$ & $16\times10^{9}\ s^{-1}$ & $\epsilon_{i}$ & $5.0\times10^{-7}$ \\ \hline
$\tau_{1}=\tau_{2}$ & $40$\ ns &$\eta$ & $2.5\times10^{8}\ s^{-1}$ \\ \hline
\end{tabular}
\caption{Parameter values used in numerical simulations.}
\label{parameters}
\end{table}
The numerical solutions were obtained with a standard fourth order Runge-Kutta algorithm. 
With the parameters used, the fastest time scale was set by $\tau_p \sim $ 2 ps and 
integration time steps were fixed at $dt = 0.2$ ps. 
Transients spanning one hundred external cavity-feedback times were discarded in the solutions.
Robustness of the results with respect to small parameter variations
was verified. Comparisons between theory and experiment are presented next.

\section{Experimental and theoretical results}

\begin{figure}[!hbtp]
 \centering
 \subfigure{
	 \includegraphics[width=4.0cm,height=4cm]{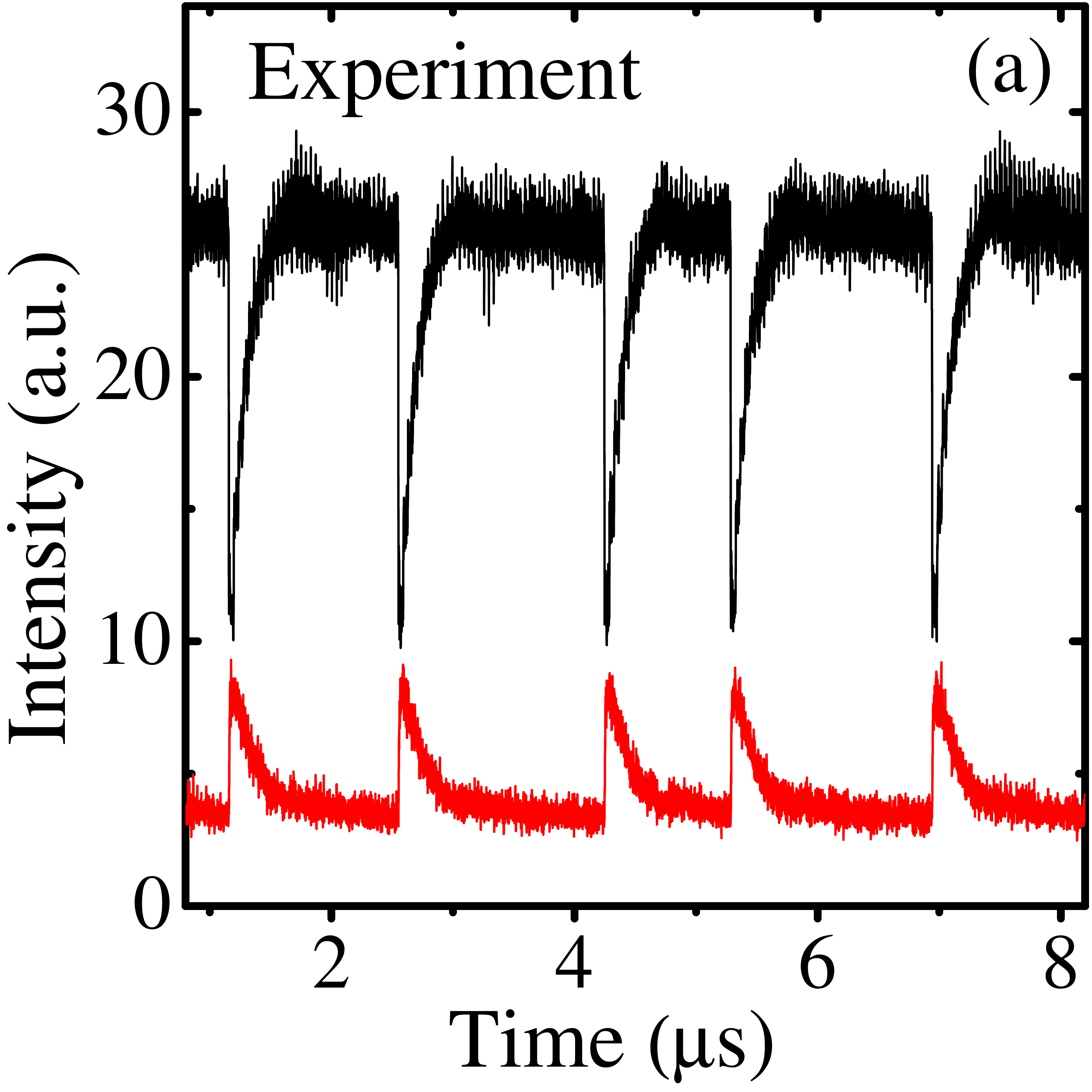}
   \label{fig:Fig2a}
	 }
\subfigure{
\includegraphics[width=4cm,height=4cm]{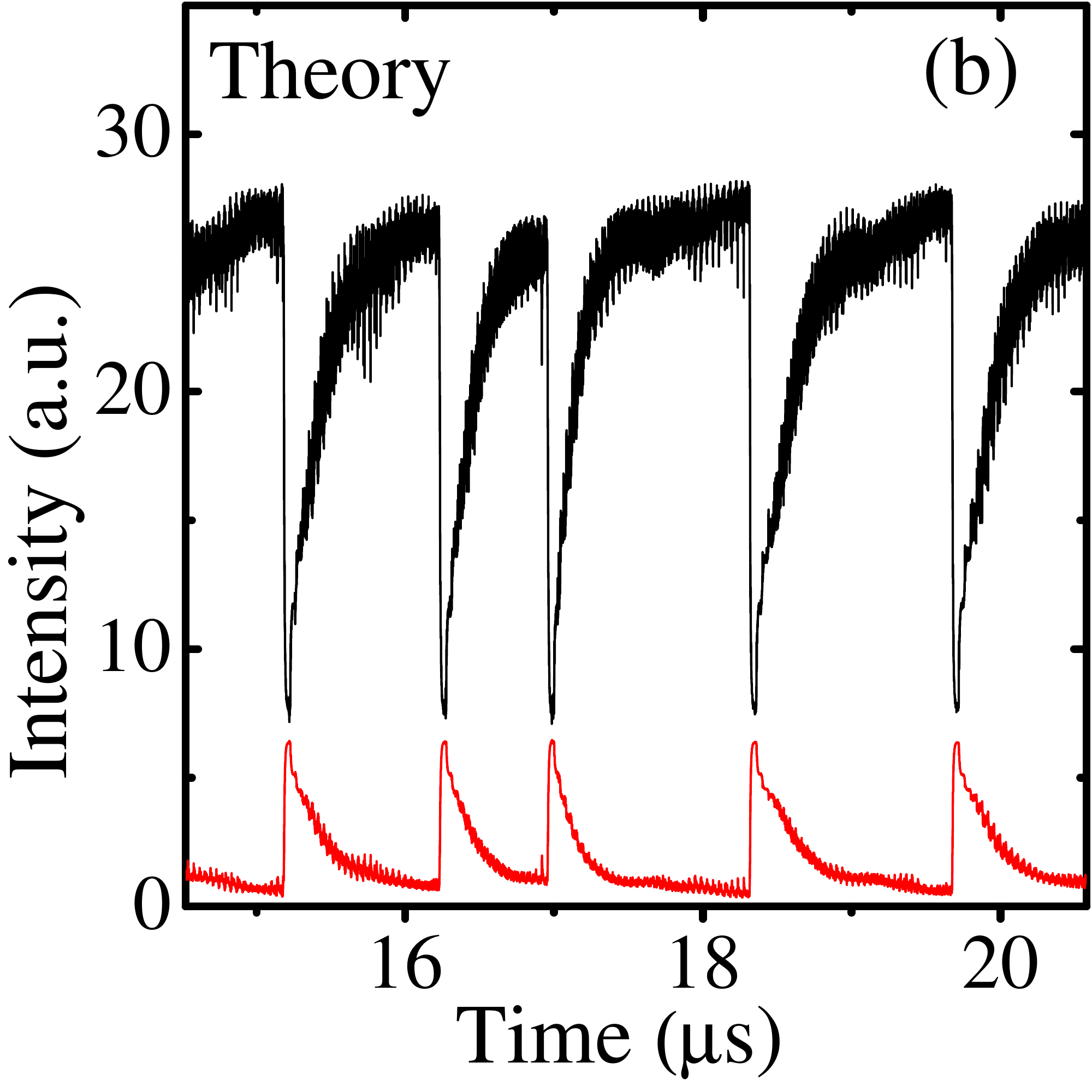}
\label{fig:Fig2b}}	  
\caption{Output power of the  coupled lasers. Laser $1$ (top line) that has optical feedback 
 shows LFF power drops while Laser $2$ (lower line) had no optical feedback.  (a) Experimental light intensity and
(b) $\left|E_j(t)\right|^{2}$ from numerical integration of equations \ref{eq:LKeq1}-\ref{eq:LKeq3}
(The vertical scales were displaced for better visualization).}
\label{fig:Fig2}
\end{figure}

Let us now give the experimental results along with the theoretical numerical integration of our equations.
Segments of the lasers power when only one laser has optical feedback are presented in Fig. \ref{fig:Fig2a}. 
The signal corresponding to Laser $1$, operating with optical feedback,
is given in the top lines (displaced for easier visualization).  
The chaotic nature of the dynamics in Laser 1 appears in the irregular time interval between LFF drops. 
Laser $2$ (lower line) was electrically coupled but did not have optical feedback. 
It presents jump up spikes in optical power, acting as a kind of sensor for the chaos in Laser 1 
via their electronic dynamics. 
Fig. \ref{fig:Fig2b} shows equivalent segments of calculated time series. 
The value of $\eta$ for the equations was extracted from the experimental signal by matching 
the relative amplitude for the drop and jump up. It corresponds to a current partition deviation from $J_0/2$ 
of $\delta J_i \thickapprox\pm 10^{-3} J_0$ at the spikes of LFFs and jump ups. 
Direct measurements of the currents fluctuations are given in section \ref{electrical}.

%

A single laser, pumped by a constant current, has less carrier recombination in its junction region 
when its power output  decrease. 
This means an increase of junction voltage, as observed by  Ray {\it et al.} \cite{roy2006}.
The number of junction carriers increases and the physical picture representing 
Current-Voltage in a direct polarized ideal diode under feedback light
gives consistent explanation to these results. 
Here, with the parallel coupled lasers, the total current is constant but the current in each laser can vary. 
The consequence is observed in Fig. \ref{fig:Fig2}. 
Each time Laser $1$ has an optical power drop, there is a decrease of its current and the correlated 
increase in the current of Laser 2 implies a jump up of its power (lower line).   
The ultrafast fluctuations (tens of picoseconds) 
are not observed in our experimental conditions.
Our detection is not sensitive at the ultra high frequencies.

Drops in light emission from one diode source accompanied by an emission increase   
from another one electrically connected in parallel were reported 
a long time ago for LEDs \cite{Edwards92} and lasers 
\cite{ElectronCoupleMIT-APL}. 
These were experiments in fluctuations around stable operation 
conditions different from the emphasis of our work.

\begin{figure}[!hbtp]
 \centering
\subfigure{
   \includegraphics[width=4.0cm,height=4cm]{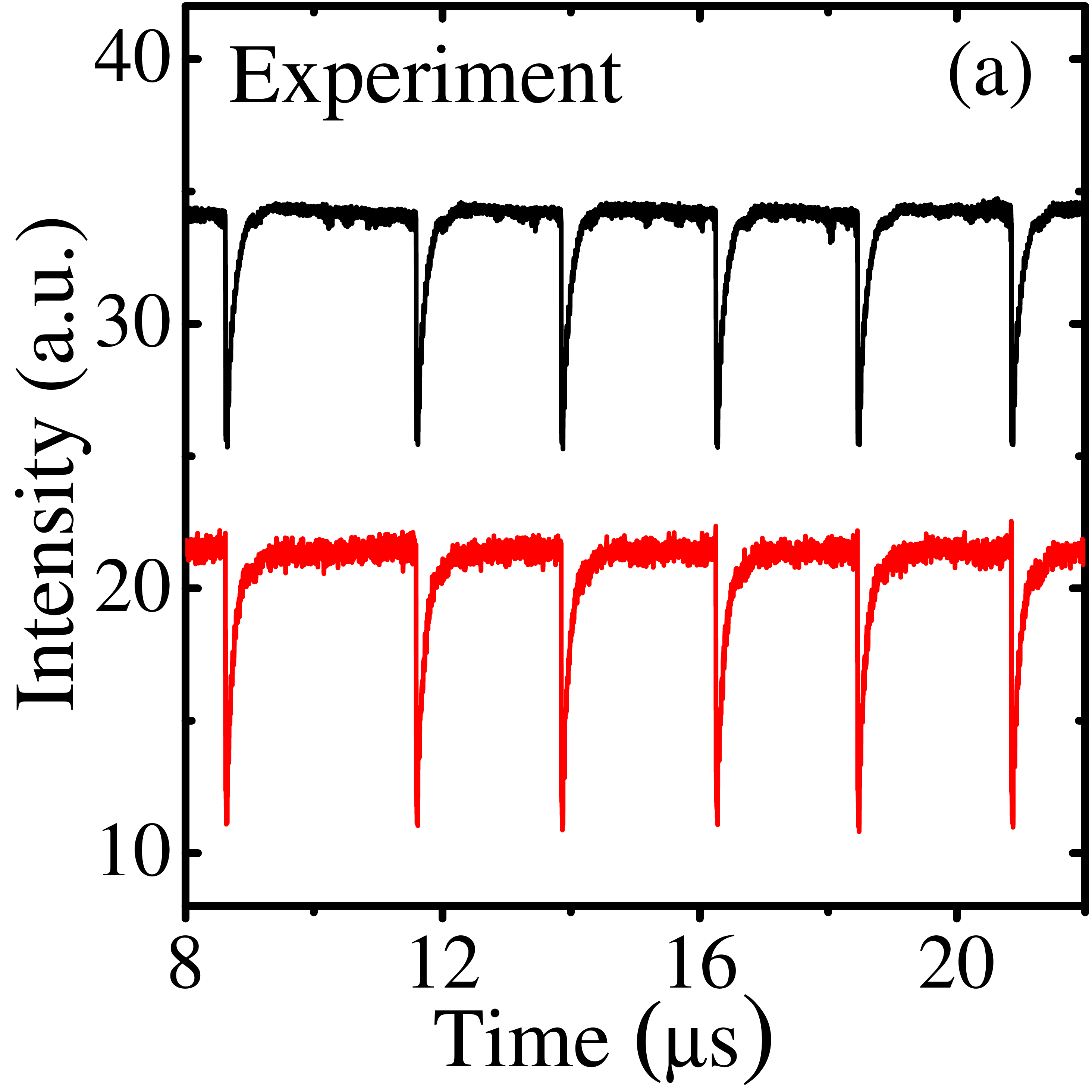} 
   \label{fig:Fig3a} } 
\subfigure{
 \includegraphics[width=4.0cm,height=4cm]{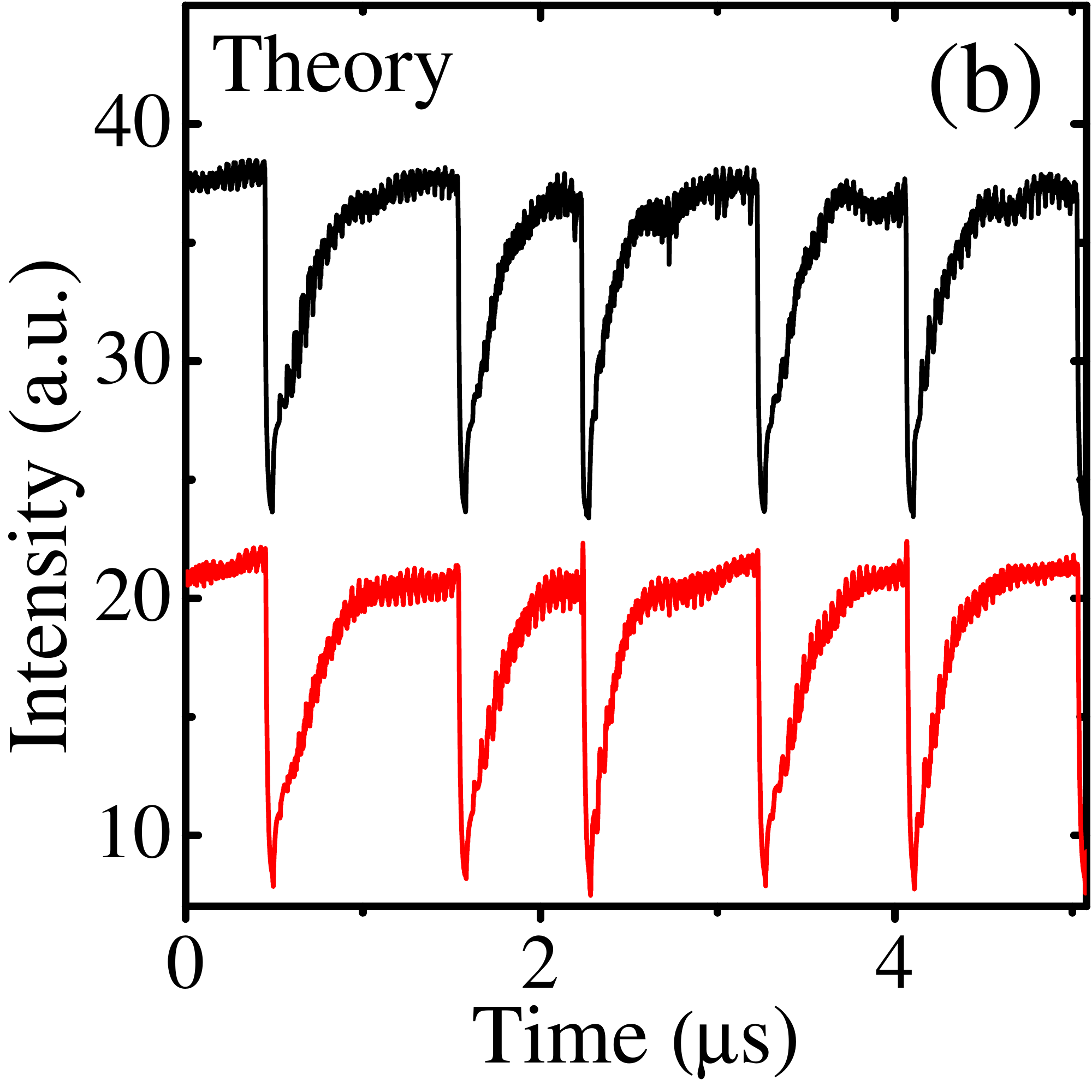}
 \label{fig:Fig3b}}
\caption{Output power of the lasers coupled in parallel and having optical feedback. 
Top line is Laser 1 and bottom Laser 2. (Power scales were displaced for visualization). 
Synchronism is clearly observed.
(a) Experimental time series including  a few power drops.  
(b) Numerical integration for the coupled laser equations.}
\end{figure}

Chaotic dynamics in the parallel electronic coupling scheme 
shows more than simply anti phase correlations.
Novel results are revealed when both lasers have optical feedback. 
In general each laser can manifest uncorrelated 
LFF power drops. However, as seen in Fig. \ref{fig:Fig3a}, when we choose appropriate 
experimental alignments and value for the total current, the large power drops
synchronize in phase. 
%
Now, instead of an increase of one 
laser power, at the expenses of the big drop in the other one, both lasers have nearly simultaneous LFF power drops.  
Figure \ref{fig:Fig3b} shows a comparable numerical segment of the lasers power 
calculated using the coupling parameter $\eta$ obtained from the data with a single laser having the 
feedback, as in Fig. \ref{fig:Fig2a}. 
The small anti-phase fluctuations are always present in any of the dynamical conditions, 
independent of LFF synchronization.  

\begin{figure}[!hbtp]
\centering
\subfigure{
\includegraphics[width=4cm,height=4cm]{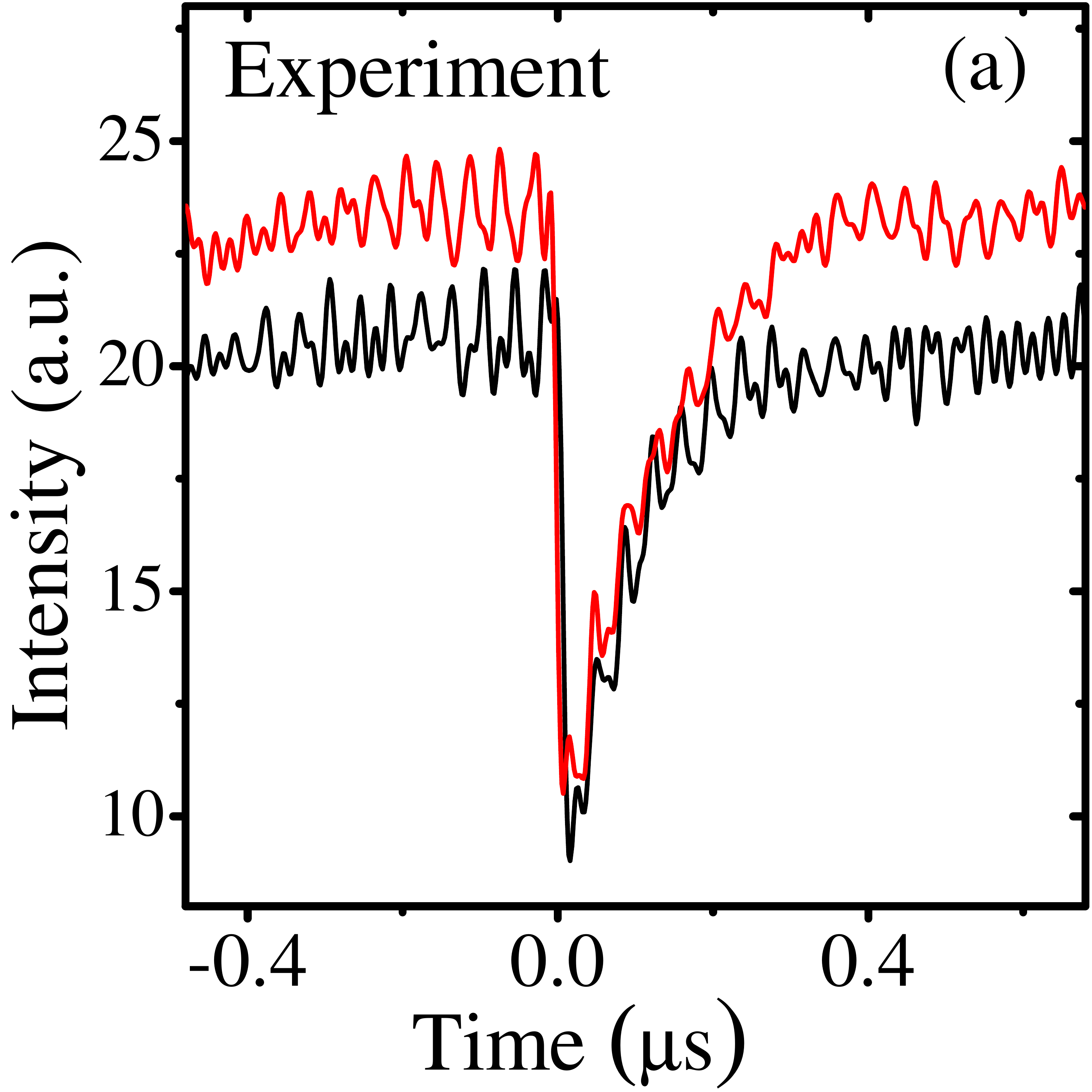}
\label{fig:Fig4a}}
\subfigure{
\includegraphics[width=4cm,height=4cm]{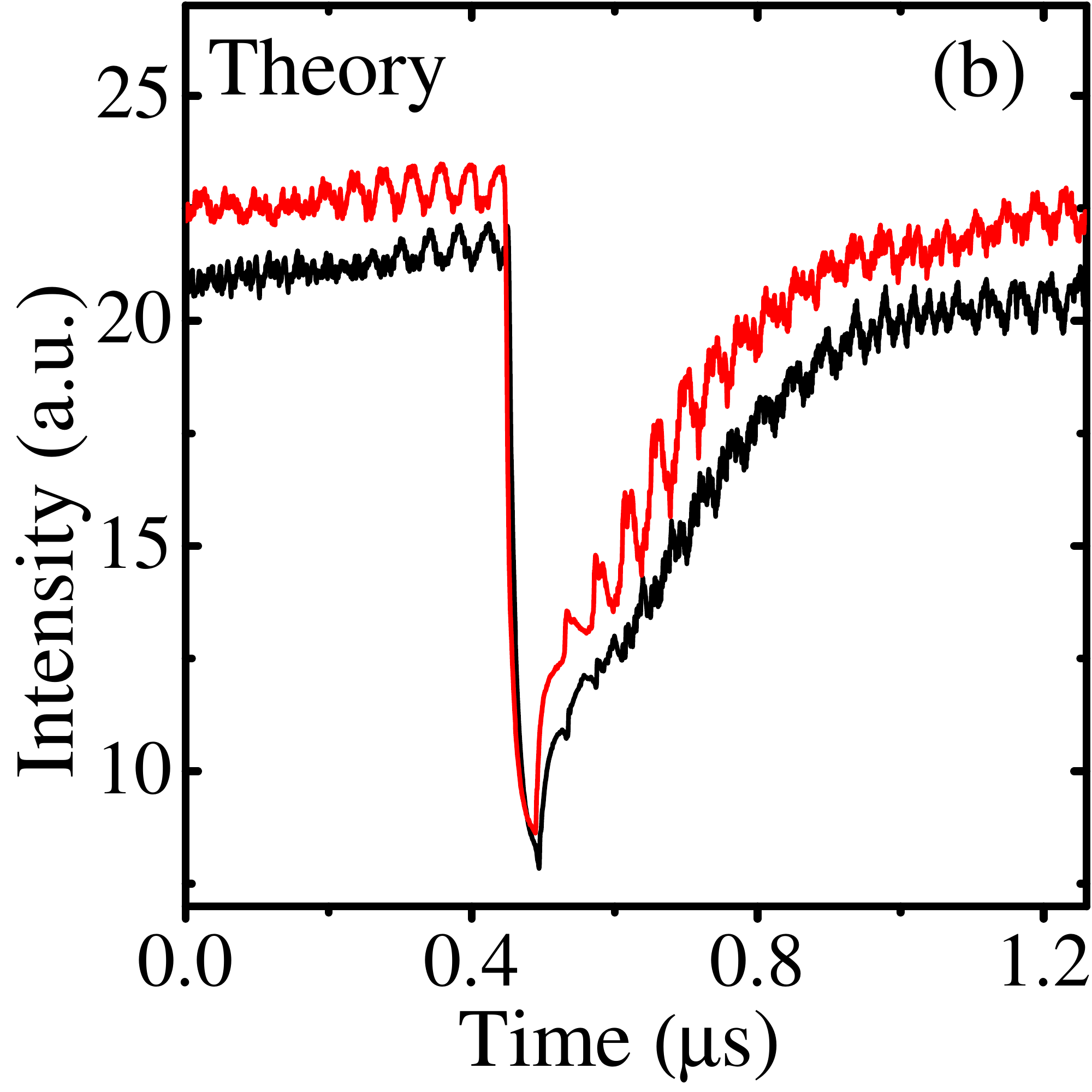}
\label{fig:Fig4b}}
\caption{ Signal showing anti phase fine structure fluctuations superimposed on a pair of drops of the in phase synchronized LFF 
when $\kappa_1=\kappa_2$. (a) Experiment and
(b) Theory. }
\label{fig:Fig4}
\end{figure}

These anti phase fluctuations at time scale above nanosecond are clearly observed as 
we look into short segment of the time series as shown in Fig. \ref{fig:Fig4}.  
They were also measured directly in the voltages and currents variation on each laser
(see section \ref{electrical}).

%

\begin{figure}[!hbtp]
 \centering
 \subfigure{
 \includegraphics[width=4cm,height=4cm]{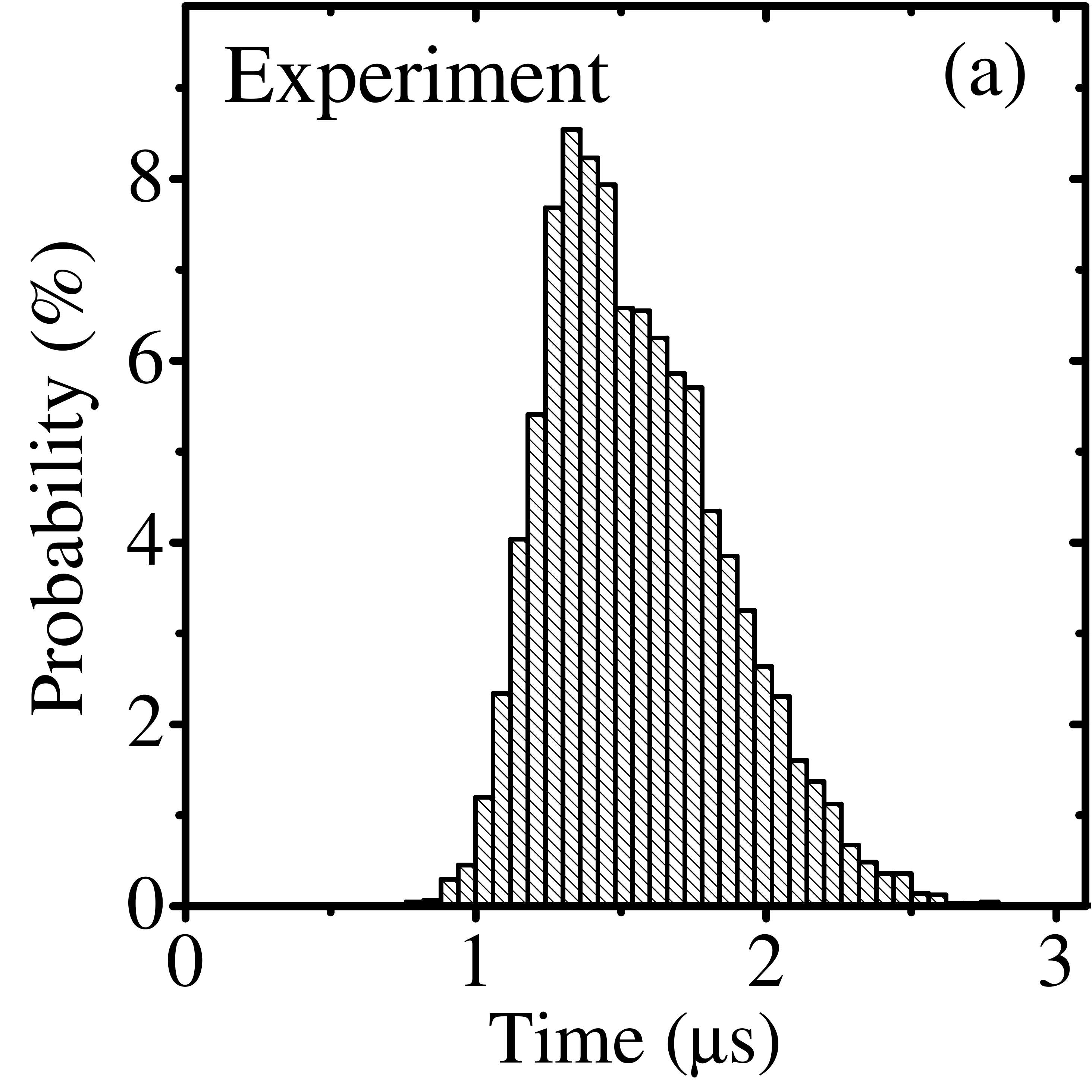}
 \label{fig:Fig5a}}
 \subfigure{
\includegraphics[width=4cm,height=4cm]{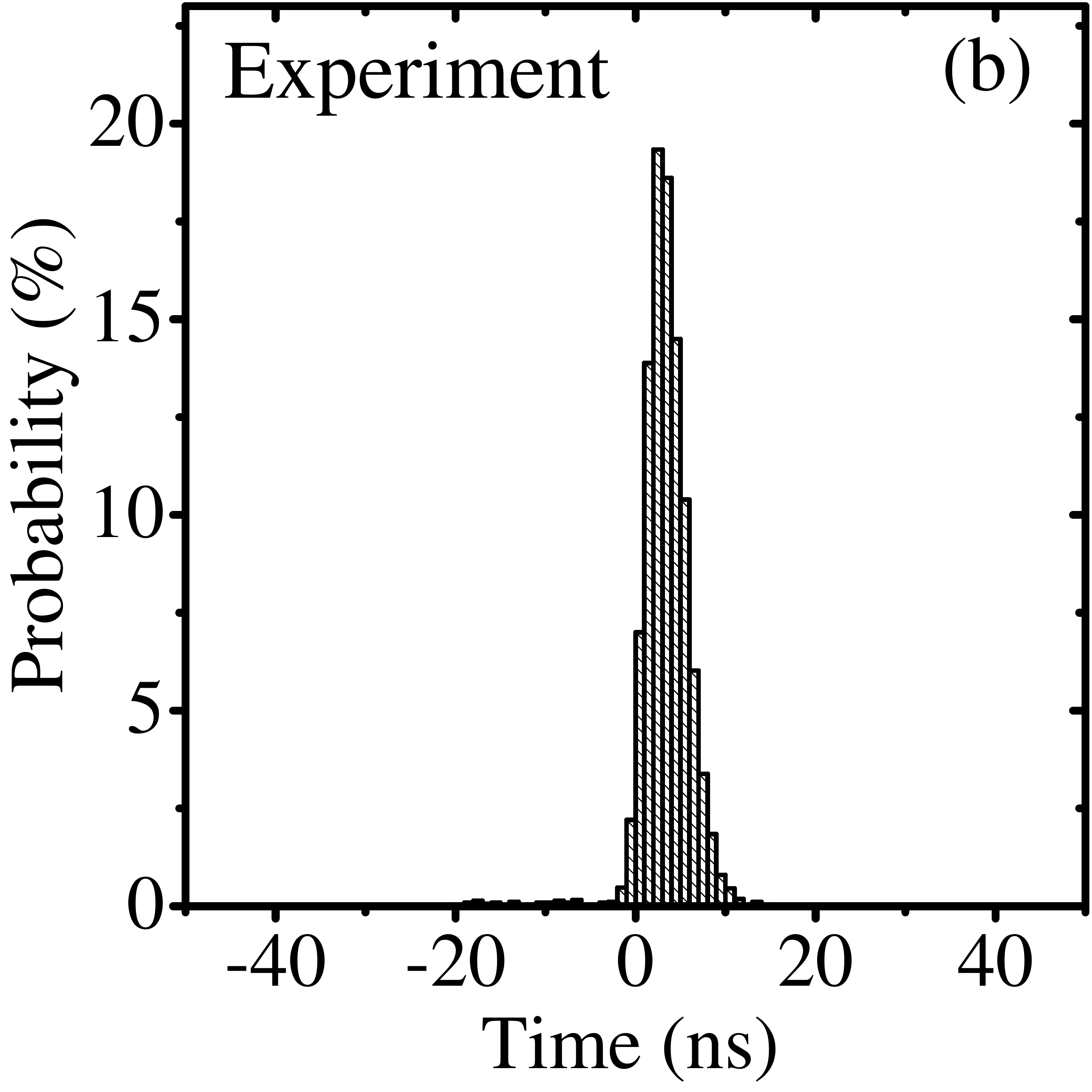} 
\label{fig:Fig5b}}
 \subfigure{
 \includegraphics[width=4cm,height=4cm]{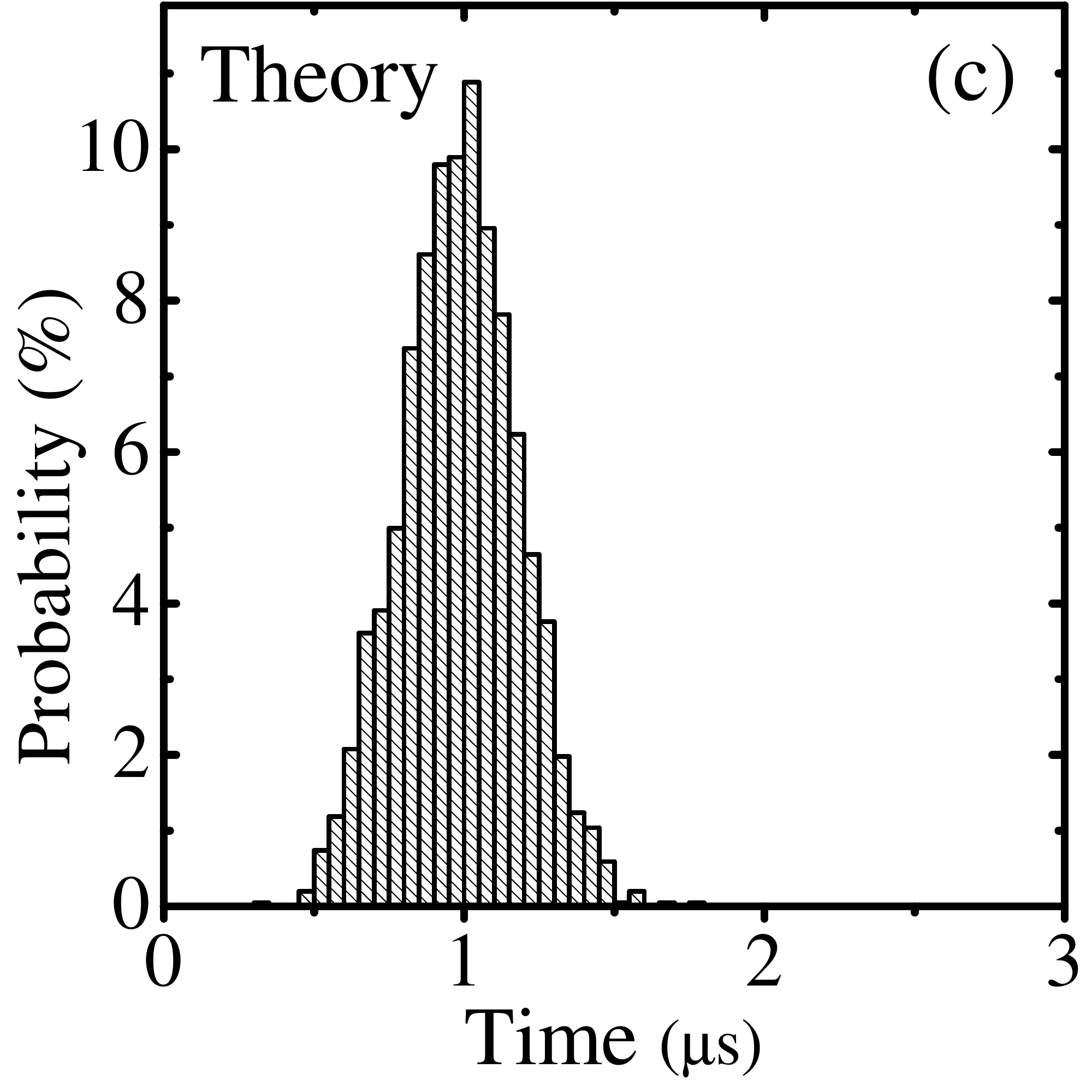}
 \label{fig:Fig5c}}
\subfigure{
\includegraphics[width=4cm,height=4cm]{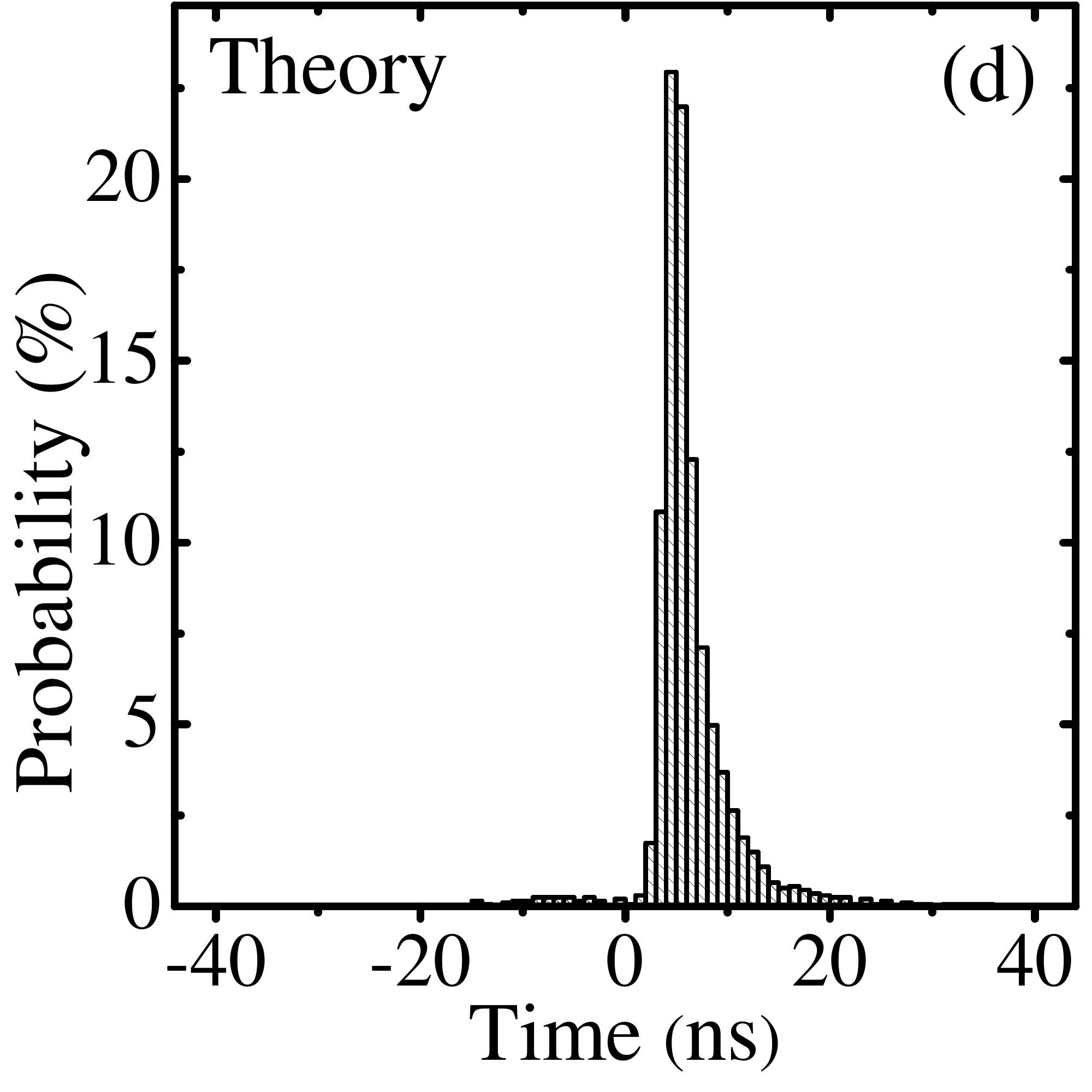}
\label{fig:Fig5d}} 
 \caption{Experimental LFF time interval histograms: (a) Time interval between successive 
 drops in one of the lasers.
(b) Time interval between drops of the two lasers. (c) Theory for (a). (d) Theory for (b)}
  \end{figure}

To better characterize the dynamics of the coupled system, histograms of events and correlation functions
were made from experimental data series. 
Within full synchronism typical data series containing more than $10^5$ events are captured without unpaired drops.
The chaotic nature 
of the LFF drops shows in the broad histograms for the time interval between consecutive drops in any one of the lasers.  
These distributions, shown in Figs. \ref{fig:Fig5a} and \ref{fig:Fig5c}, approximate  Gaussians with an average 
time between drops of $1.5 \mu$s, which is 37 times the feedback time, and a wide
variance close to $1 \mu$s. Both quantities are more than one order of magnitude larger than the lasers feedback time. 
The synchronization indicator is represented in the histograms given in Figs. \ref{fig:Fig5b} and \ref{fig:Fig5d}
for the time difference between the power drops of the two lasers. 
It has a very narrow distribution, centered near zero delay. 
%

Notice in Fig. \ref{fig:Fig5b} that almost all pairs of drops occur within less than $10$ ns, which 
is significantly shorter than the experimental $40$ ns of optical feedback time. The small shift of a few nanoseconds
from zero delay is sensitive to experimental unbalance of laser parameters.  
When one laser  
has higher value for its feedback coefficient and/or lower threshold 
and/or higher pump current in the parallel coupling circuit  it becomes leading in the drops.
The effects of small unbalances on the synchronism shifts were confirmed in the 
theoretical-numerical solutions as shown in Fig. \ref{fig:Fig5d}. 
Bimodal distributions, with symmetrical anticipations and delays in the synchronizations, were also observed in 
experimental data and obtained in the theory, as we properly bias the parameters. 
 
\begin{figure}[!hbtp]
\centering
\includegraphics[width=7.7cm,height=6.05cm]{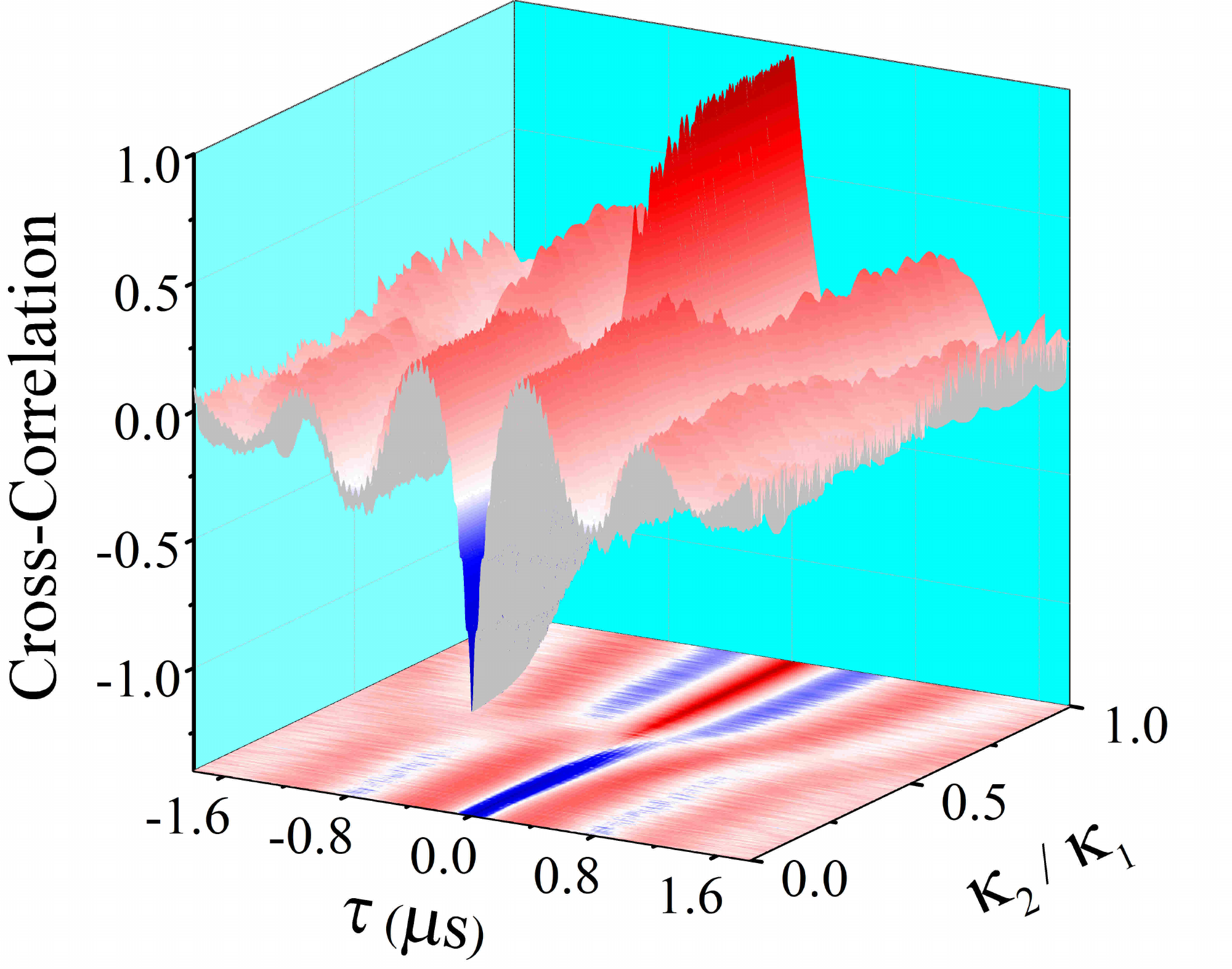}
\caption{Theoretical numerical cross-correlation dependence on the feedback coefficient ratio.}
\label{fig:Fig6}
\end{figure}

\begin{figure}[!hbtp]
 \centering
 \subfigure{
 \includegraphics[width=4cm,height=4cm]{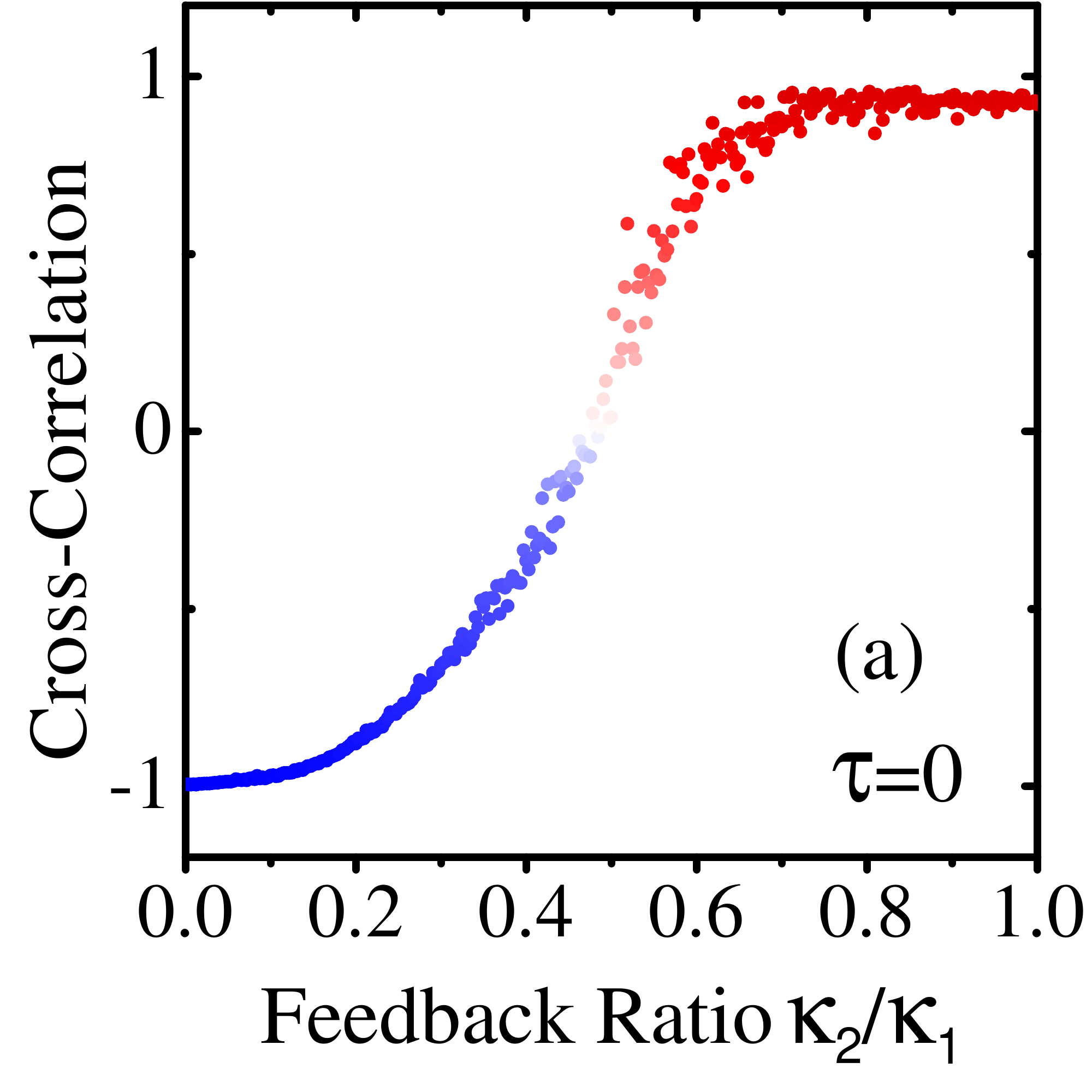}
  \label{fig:Fig7a}}
  \subfigure{
  \includegraphics[width=4cm,height=4cm]{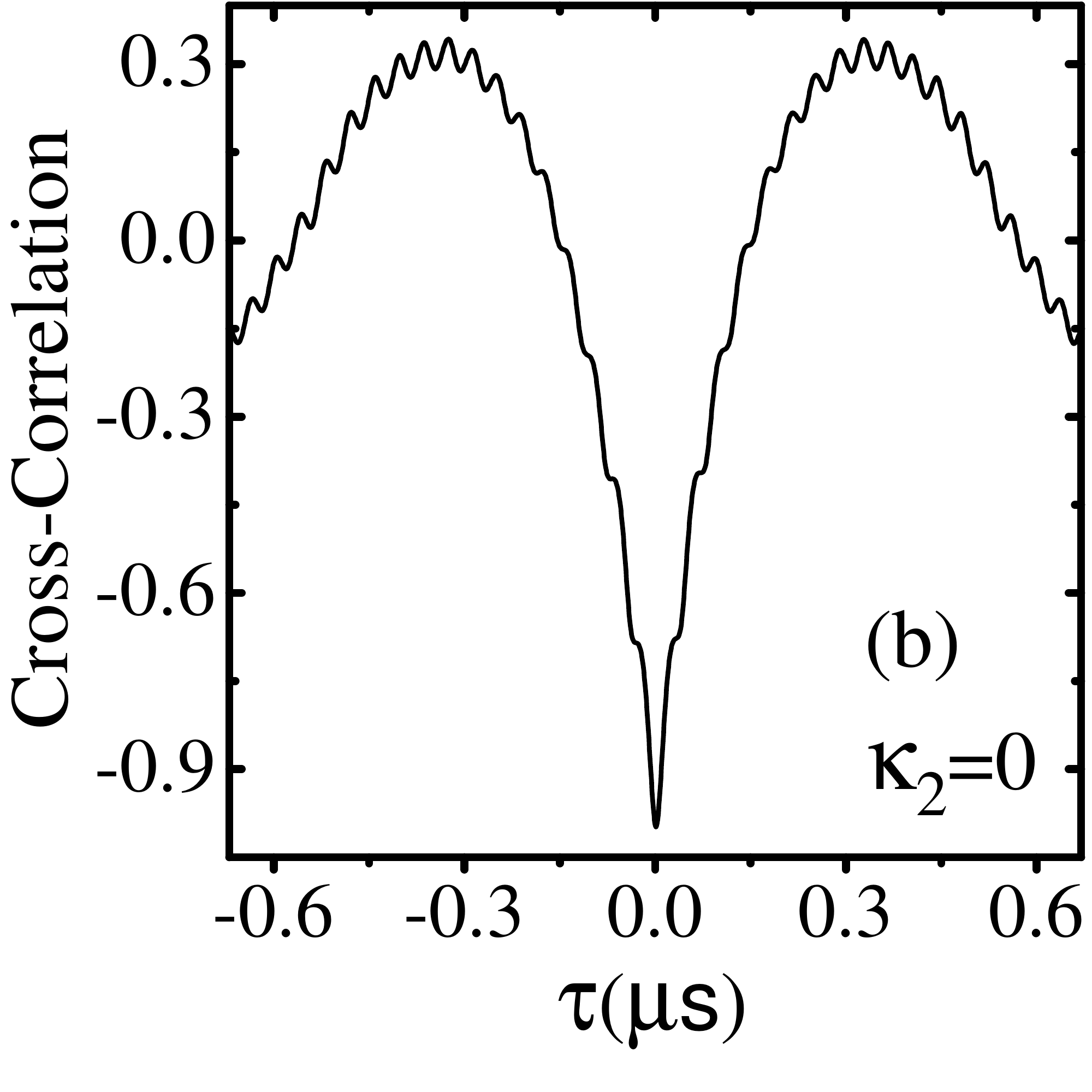}
 \label{fig:Fig7b}}
 \caption{Detail of Fig. \ref{fig:Fig6}. (a) Cross correlation between the lasers calculated at 
 	zero delay time as a function of $\kappa_2/\kappa_1$. (b) 
 Cross correlation between the lasers calculated when Laser 2 has no feedback, ($\kappa_2=0$).}
\label{fig:Fig7}
  \end{figure}

\begin{figure}[!hbtp]
\centering
\includegraphics[width=7.7cm,height=6.05cm]{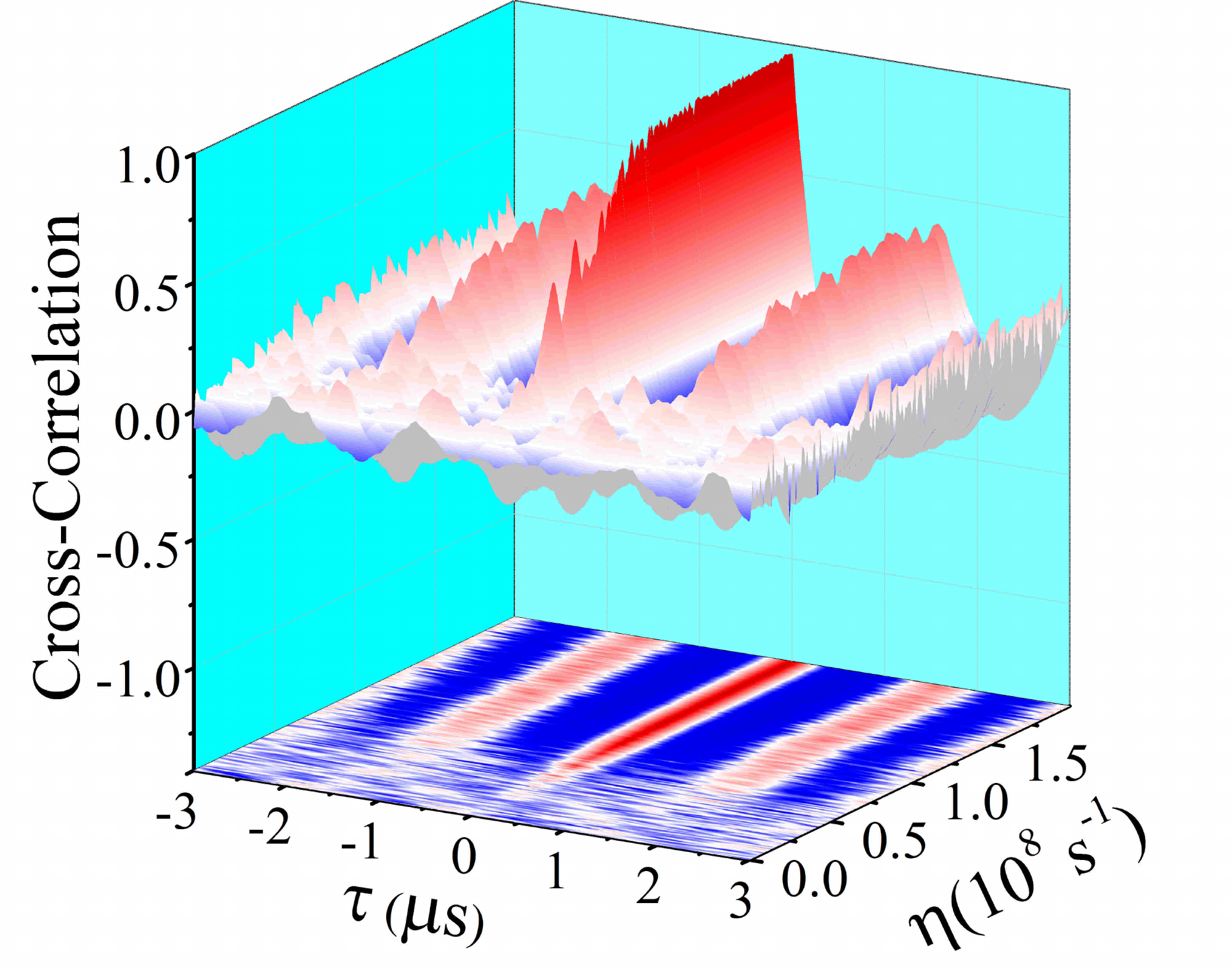}
\caption{Theoretical numerical cross-correlation dependence on the coupling coefficient $\eta\,$. 
Same parameters from the Table \ref{parameters} but $N_{02}=0.998\times10^{8}$ and $J_0=2.02J_{th1}$.}
\label{fig:Fig8}
\end{figure}

The onset of LFF synchronism depends on the feedback coefficient in each laser and on their coupling 
impedance. We present next how the synchronism transition, 
calculated as a function of the 
parameters $\kappa_2/\kappa_1$ 
and $\eta$, manifest in the cross correlation functions $C(\tau)$ shown in Figs.\ref{fig:Fig6}-\ref{fig:Fig8}.  
The calculation with constant coupling parameter had $\eta$ from table \ref{parameters}.
Laser 2 is assumed to increase its 
feedback coefficient $\kappa_2$ from zero to $\kappa_2=\kappa_1= 16\times10^{9}\ s^{-1}$.  
Figure \ref{fig:Fig6} shows the calculated cross correlation function. For zero and very small optical feedback  
in Laser 2, when no LFF exists in this laser, only anti phase fast oscillation and the jump ups, 
are present, as in Fig. \ref{fig:Fig2}.
This gives the value of $-1$ for the zero delay, $\tau=0$, cross correlation.
As $\kappa_2$ increases, LFF drops start to appear in Laser 2 and some drops coincide with those in Laser 1. 
Positive contributions begin to add to the cross correlation. At some intermediate value of $\kappa_2$,  $C(\tau=0)$ vanishes.
This does not mean that the signals are totally uncorrelated. It does show that in-phase and anti-phase 
fluctuations, in different time scales, are simultaneously present in the dynamics.
The full chaos synchronized LFF dynamics only appears with $\kappa_2/\kappa_1 > 0.7$ and 
 the cross correlation is almost $+1$ (there is an ever-present small anti phase contribution). 
%
Cross sections of the 3D Fig. \ref{fig:Fig6} can reveal the onset of LFF synchronization when 
we plot  $C(\tau=0)$ as a function of $\kappa_2/\kappa_1\,$. It can also 
show the anti phase fluctuations in the graph of  $C(\tau)$ for $\kappa_2=0\,$. 
These are given in Figs \ref{fig:Fig7a} and \ref{fig:Fig7b}, respectively.

%
The cross correlation function was also calculated varying the electronic coupling coefficient to 
show the onset of LFF synchronization, starting with the two lasers having independent LFFs. 
This is given in Fig. \ref{fig:Fig8}. Very small values of $\eta$
are unable to mediate the synchronism. Therefore the two lasers have  LFF drops but they are 
independent and their cross correlation is null. As the value of $\eta$ increases, a transition region is 
reached where more and more LFF pairs drop near simultaneously.
%
At the value of $\eta= 1.2\times10^{8}$ $s^{-1}$ full synchronism is attained and $C(\tau=0)$ rises to the value 
near $+1$. This value of $\eta$ corresponds to a peak coupling 
current of $50\, \mu A$ as shown in section \ref{electrical}.

\section{Electrical measurements in the Coupled Lasers}
\label{electrical}

\begin{figure}[!hbtp]
	\centering
	\includegraphics[width=7.0cm,height=4.0cm]{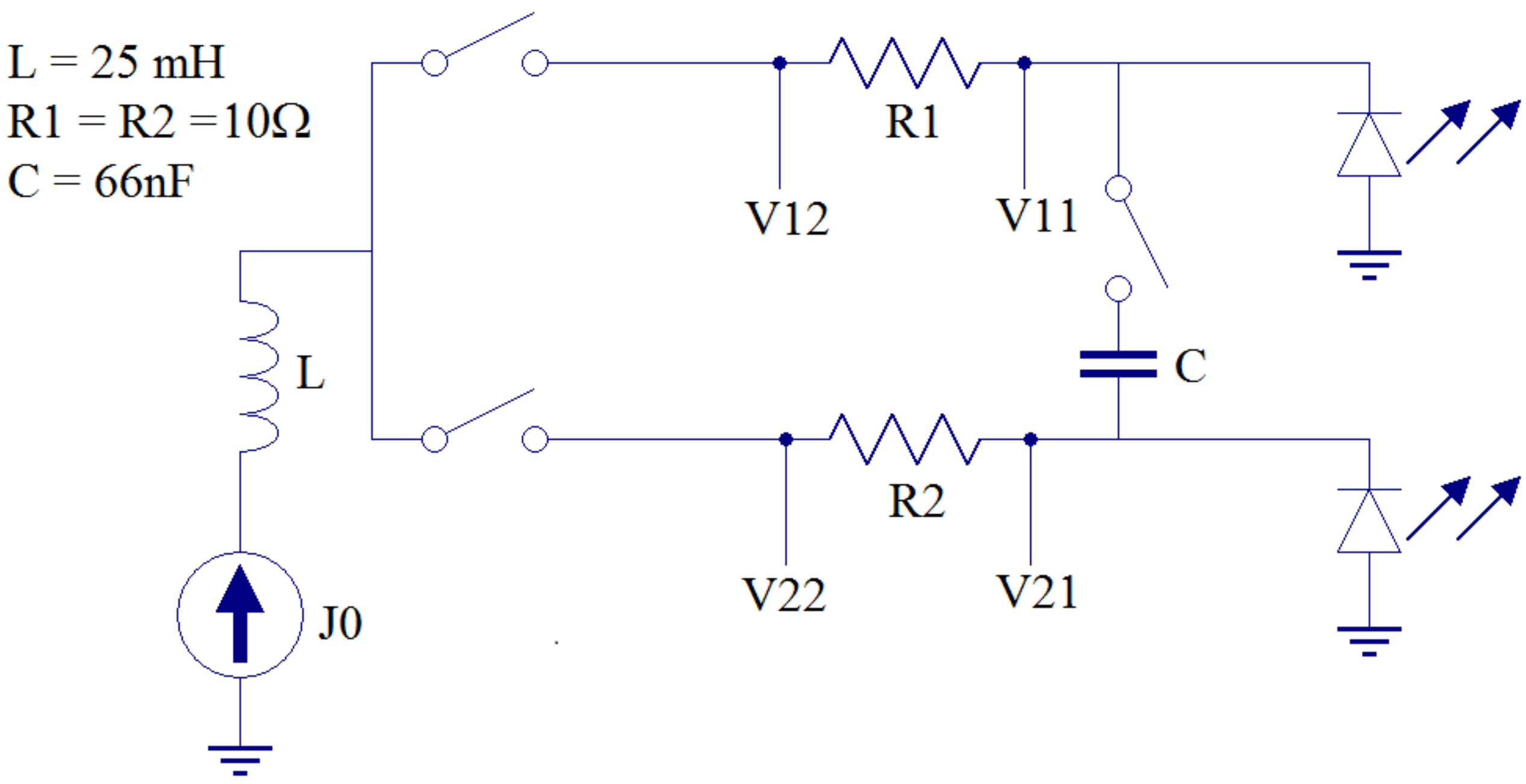}
	\caption{Setup for the experiments on the power correlations and chaos
		synchronization of two electrically coupled lasers. }
	\label{fig:Fig9}
\end{figure}

We describe here details of the electrical measurements made along with the optical measurements on the coupled lasers.
The main experiments were done with pairs of semiconductor lasers differing by less than $2 \%$ in their 
threshold current and optical 
frequency . They were coupled electrically in parallel configuration and  
pumped by a high impedance current source.
The electronic circuit had details as indicated in Fig. \ref{fig:Fig9}.
%
The  resistors $R_1$ and $R_2$ were included with two purposes. First, they controlled the 
amount of coupling between the lasers.  Their values partially determine $\eta$, the current coupling coefficient, 
through the partition of the total current by Kirchhoff\textquotesingle s laws
$J_1(t)+J_2(t)=J_0\,$, supplemented by a linearized approximation for the currents unbalance as a function of the carriers populations,
\begin{equation}
J_{1}(t) =J_{0}/2 - \eta [N_{1}(t)-N_{2}(t)]\;.
\end{equation} 
$N_i(t)$ is the active carrier population of laser $ i$ with $i=1,2$.
The internal resistances of the lasers are smaller than $5 \Omega$ \cite{ElectronCoupleMIT-APL}. The value of $\eta$
decreases with the circuit impedance between the lasers. Without the capacitor C such impedance is dominated by 
the external resistors.
Consistently, we verified that when $R_1=R_2 \geq 50\, \Omega$, 
the coupling  is reduced to the point of preventing LFF synchronism. 
With these high values, optical power measurements with just one laser having optical 
feedback, like the one reported in Fig. \ref{fig:Fig2}, did not show detectable 
response on the second laser without feedback. This  evidences small coupling and explains why
the two lasers with optical feedback had LFF but never got synchronized. 
The switch in Fig. \ref{fig:Fig9} that 
shunted the capacitor $C$ across the resistors could
restore the synchronization, demonstrating that the precise value of  $\eta$ depends on more than circuit resistors.
Our model provides a remarkable agreement with  optical and electrical measurements, as 
described below. 
The second use of the resistors was to obtain the laser currents through the voltages 
$V_{11}$,  $V_{12}$, $V_{21}$ and $V_{22}$ on the extremes of the resistors.
For that case we took small values $R_1=R_2= 10\, \Omega$ so that we had synchronism.  

\begin{figure}[!hbtp]
	\centering
	\subfigure{
		\includegraphics[width=4cm,height=4cm]{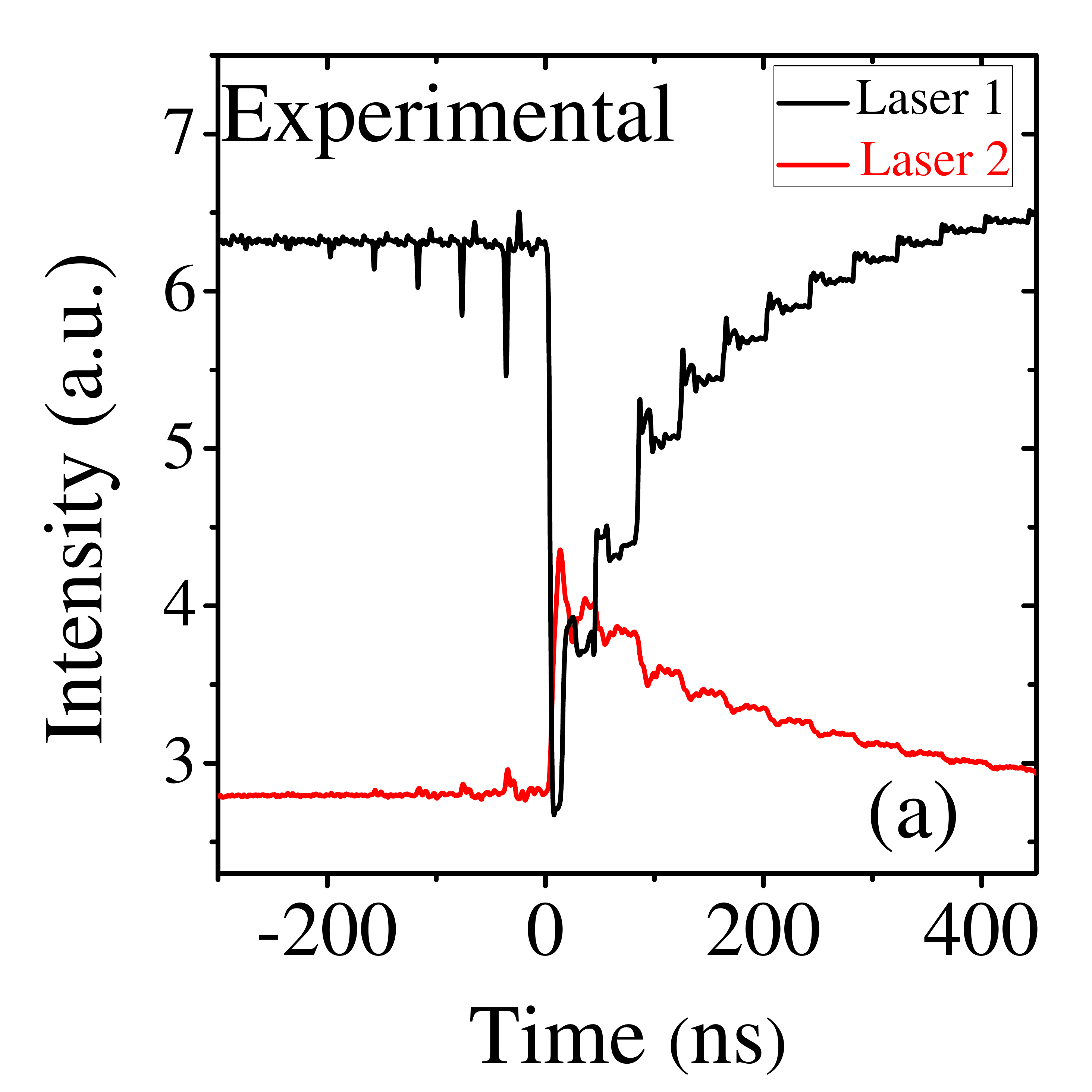}
		\label{fig:Fig10a}}
	\subfigure{
		\includegraphics[width=4cm,height=4cm]{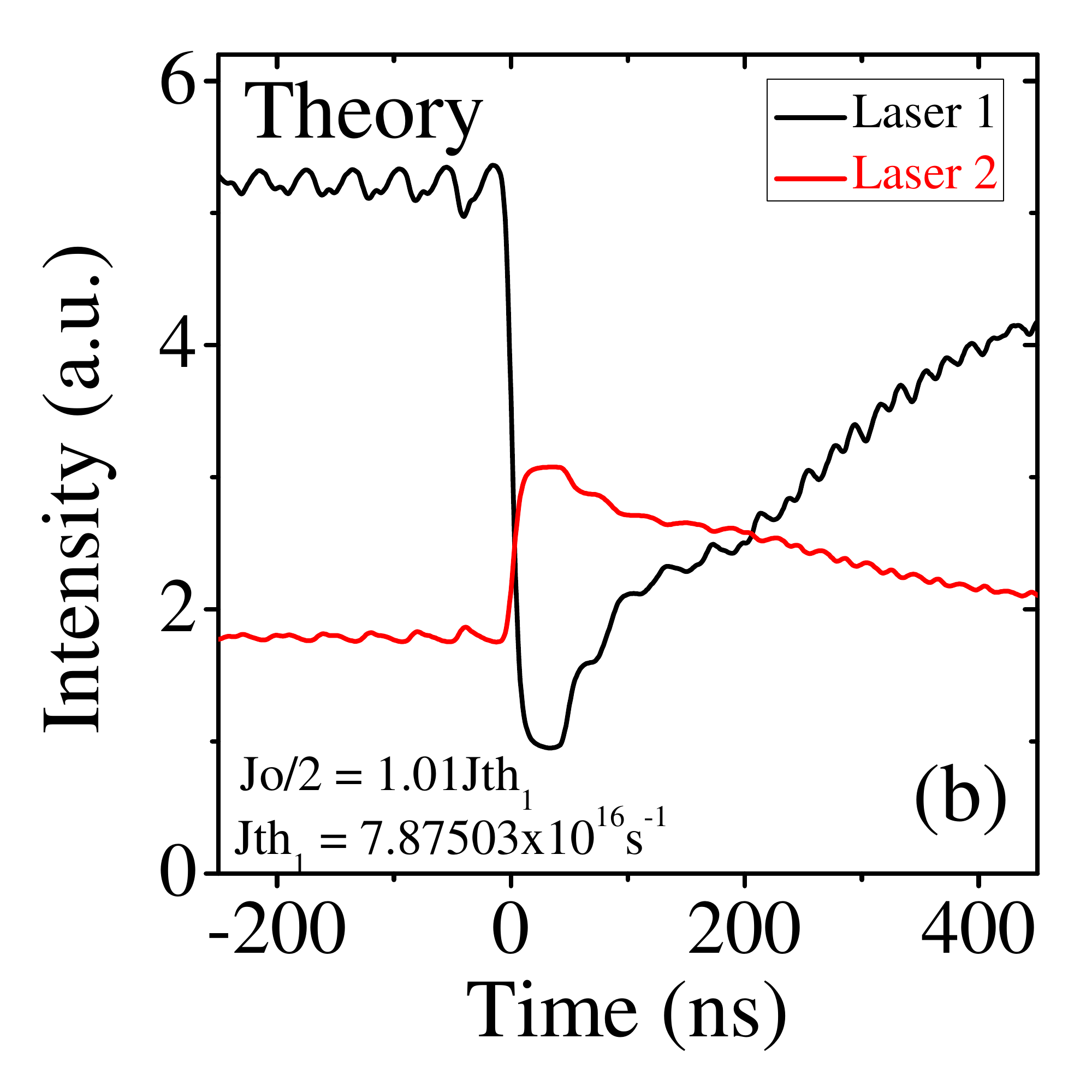}
		\label{fig:Fig10b}}
	\caption{Laser power variations during the LFF drop when just Laser 1 had feedback. (a) Experimental 
		(b) Theory. }
	\label{fig:Fig10}
\end{figure}

\begin{figure}[!hbtp]
	\centering
	\subfigure{
		\includegraphics[width=4cm,height=4cm]{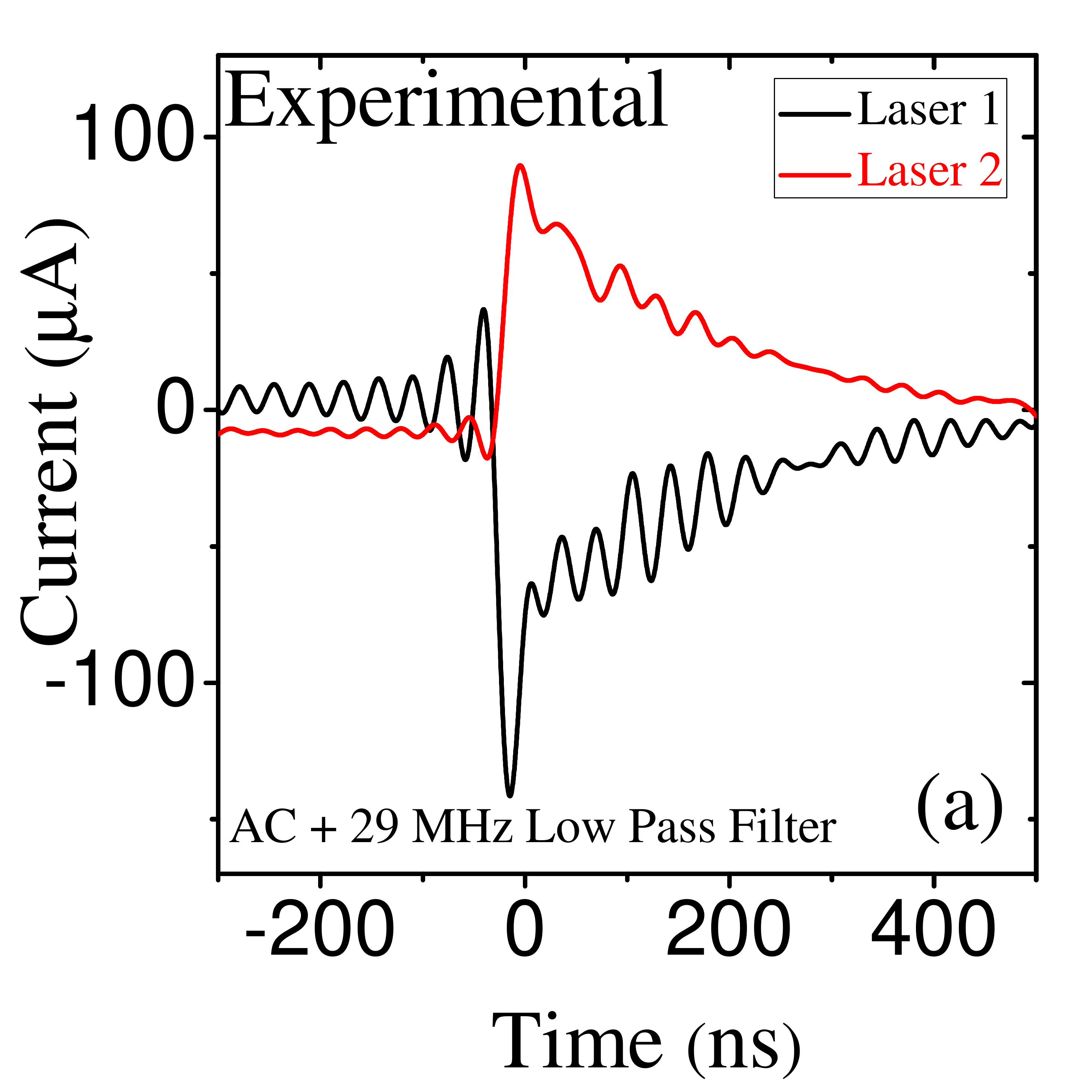}
		\label{fig:Fig11a}}
	\subfigure{
		\includegraphics[width=4cm,height=4cm]{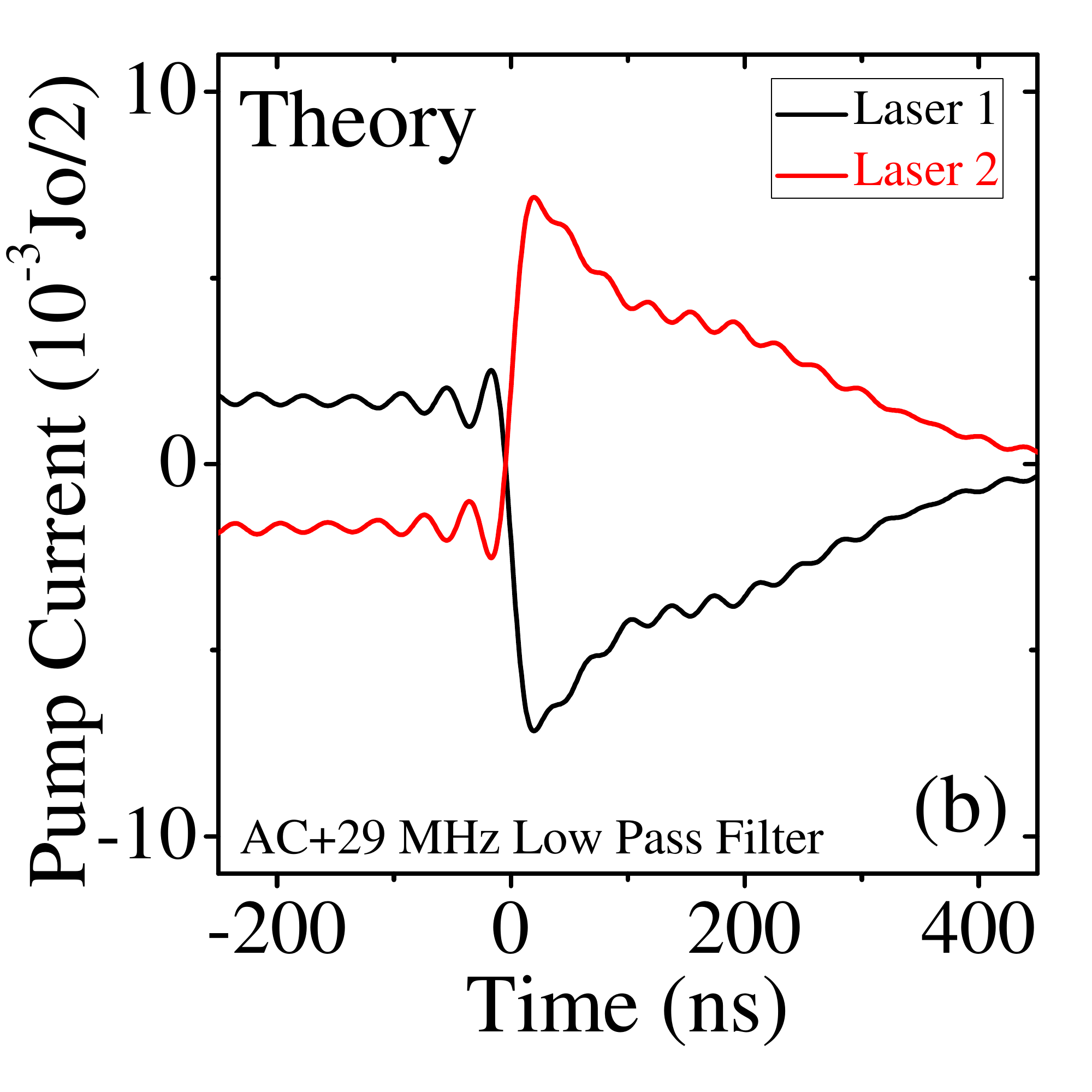}
		\label{fig:Fig11b}}
	\caption{Pump current variations during one LFF drop of laser 1 while laser 2 had no feedback: (a) Experimental 
		(b) Theory}
	\label{fig:Fig11}
\end{figure}

\begin{figure}[!hbtp]
	\centering
	\subfigure{
		\includegraphics[width=4cm,height=4cm]{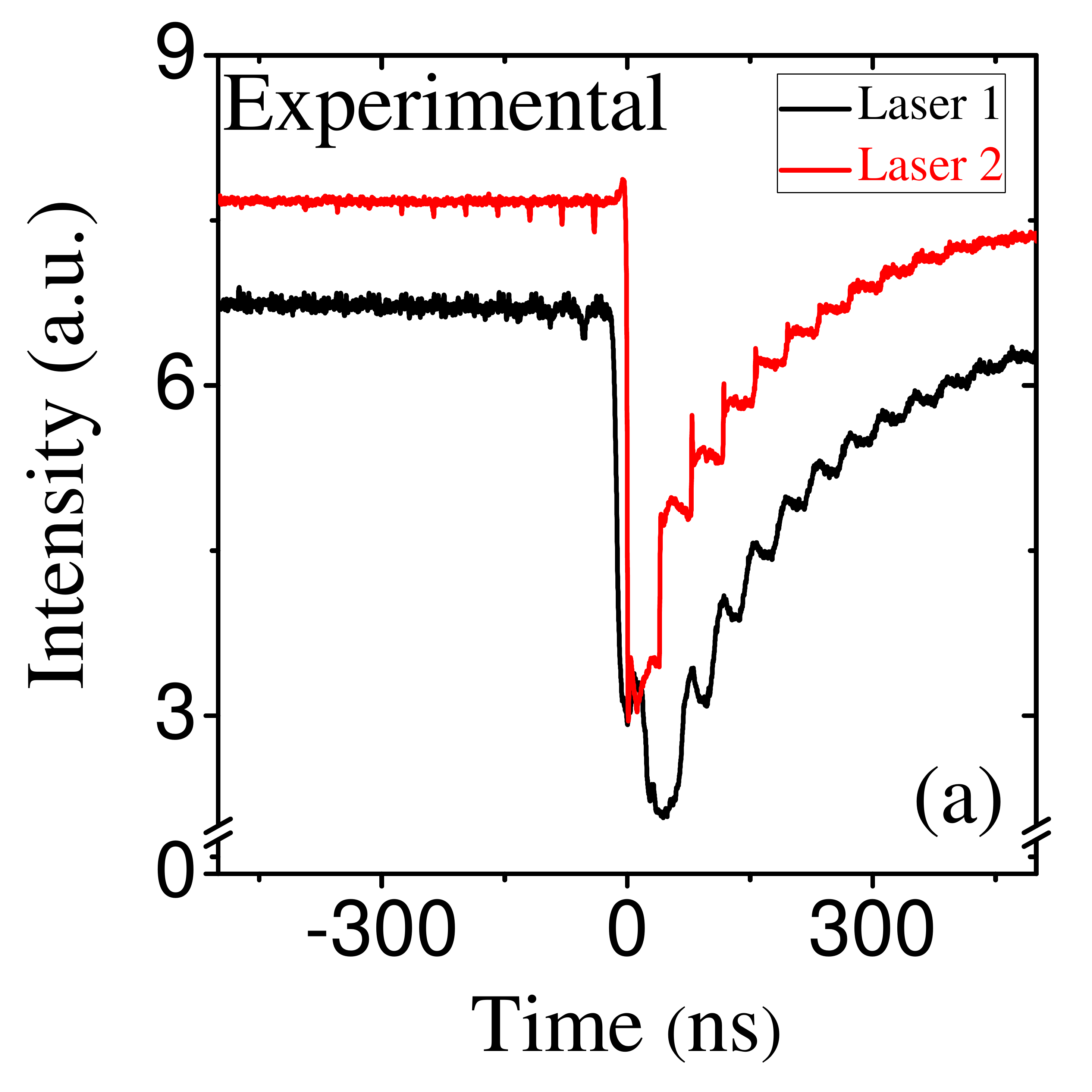}
		\label{fig:Fig12a}}
	\subfigure{
		\includegraphics[width=4cm,height=4cm]{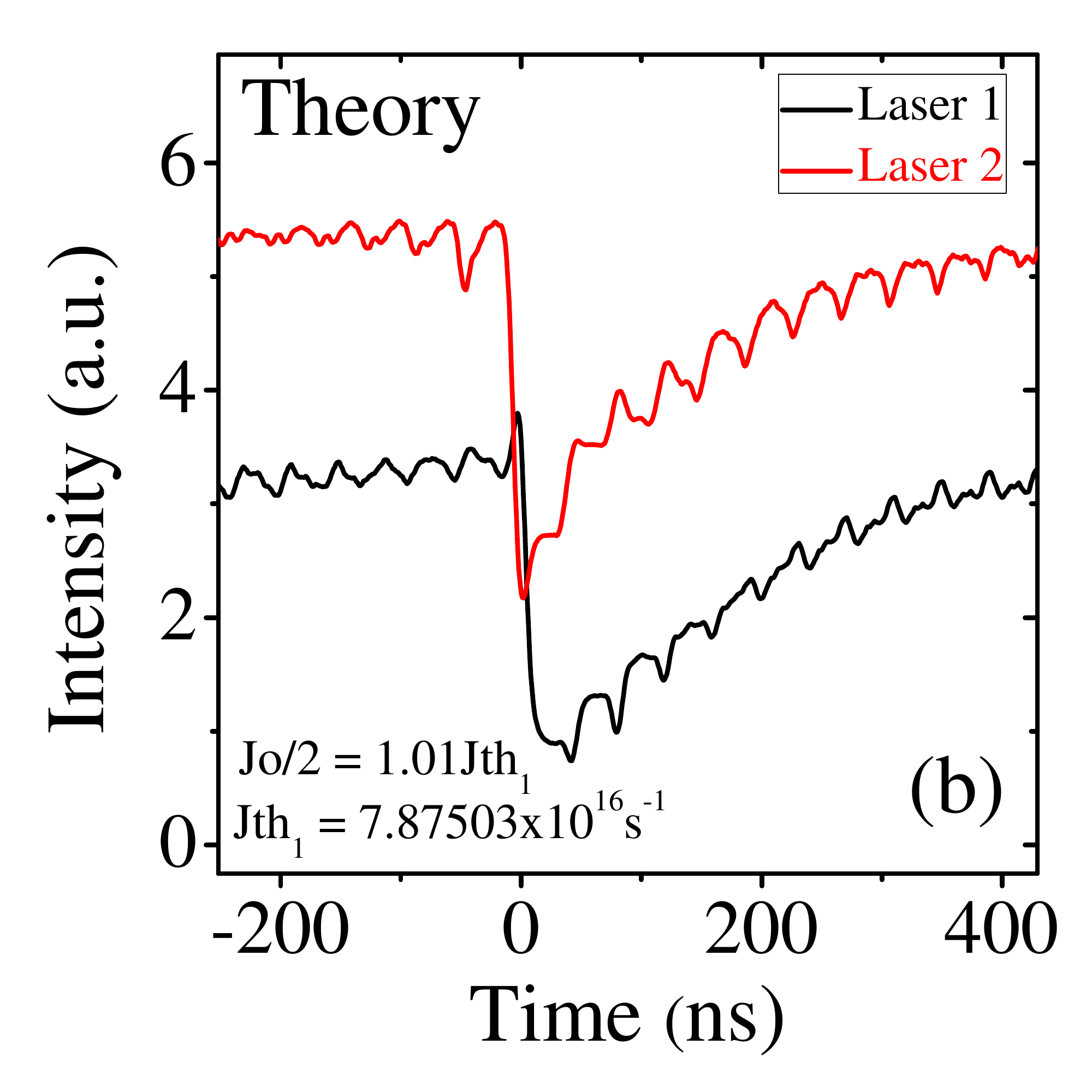}
		\label{fig:Fig12b}}
	\caption{Laser power variations during a pair of synchronized LFF drops (a) Experimental 
		(b) Theory. }
	\label{fig:Fig12}
\end{figure}

\begin{figure}[!hbtp]
	\centering
	\subfigure{
		\includegraphics[width=4cm,height=4cm]{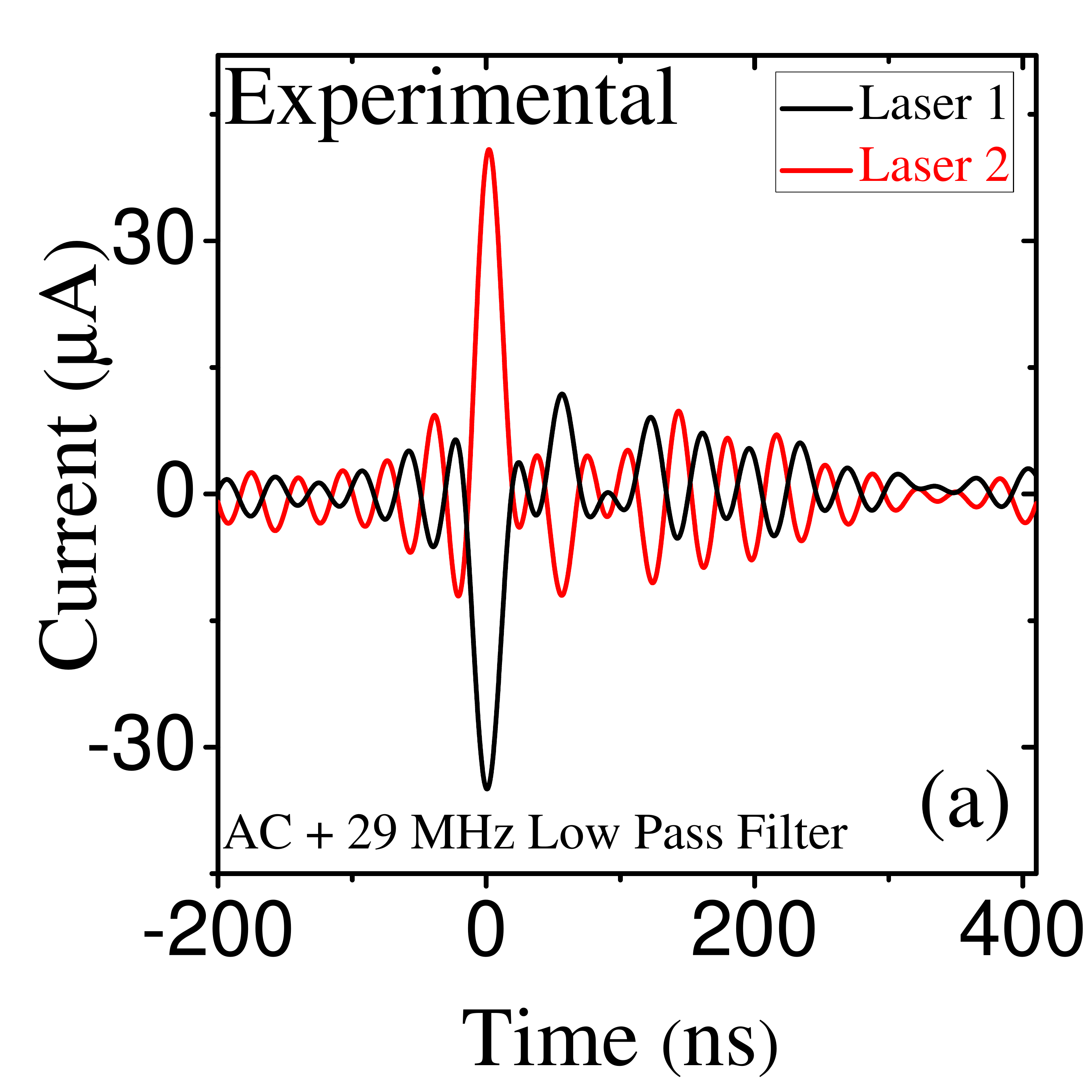}
		\label{fig:Fig13a}}
	\subfigure{
		\includegraphics[width=4cm,height=4cm]{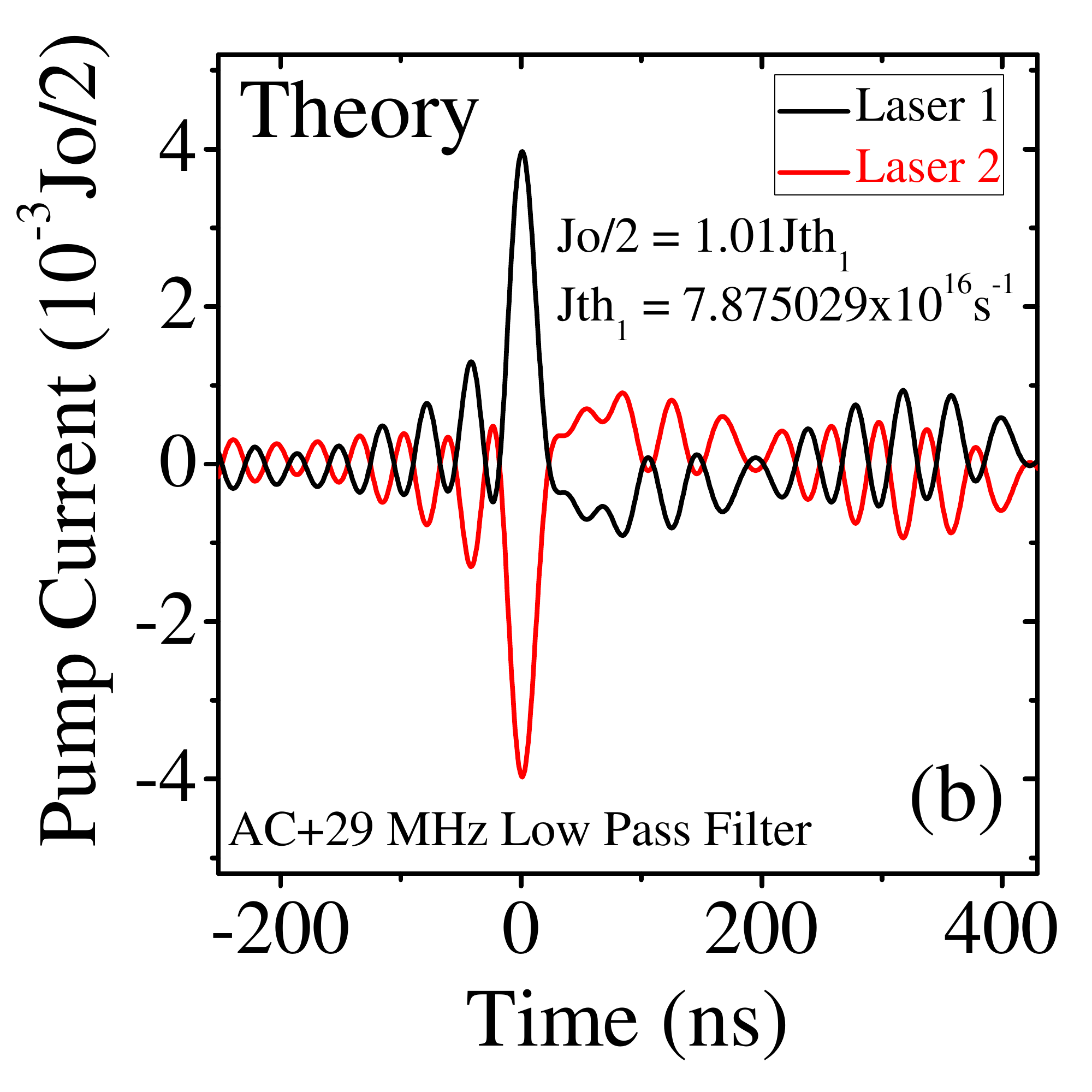}
		\label{fig:Fig13b}}
	\caption{Pump current variations during a pair of synchronized LFF drops (a) Experimental 
		(b) Theory. 
}		
	\label{fig:Fig13}
\end{figure}

We made most of our measurements on a L780P010 with threshold of $8.5$ mA emitting near 
$780$ nm wavelength and a ThorLabs L850P010 with $10.5$mA threshold current and $850$ nm
wavelength.
Optical feedback was implemented by  mirrors located at 
distances between $3.00$ and $8.00$ m from each laser, 
preceeded by beam collimation with aspheric anti-reflection coated 
lenses. The feedback field coefficients  $\kappa_i$ was determined by the 
current threshold reduction.
Up to  $6 \%$ could be achieved in both lasers.  
The feedback delay times 
had a ($\pm 5$) ps precision mismatch 
that did not affect the results. Output coupling beam splitters lead $4 \%$ of light onto 2 GHz
bandwidth photodiodes, where we captured the output power (or intensity). 
The pump current was near twice the (almost equal) single laser solitary threshold. 
Data series were acquired by fast ($>$3GHz band) photodiodes and a digital
oscilloscope having a bandwidth of $1$ GHz and a maximum sampling rate of
$5$ GS/s. The time series were computer treated to achieve averages and 
experimental histograms. Results for measurements and 
numerical calculations with the dynamical equations are shown 
in Figs. \ref{fig:Fig10} to \ref{fig:Fig13}.  
%
Notice  in Fig. \ref{fig:Fig10a} that there is a time mismatch of near 5 ns, which correspond to $1/10$ of the feedback time.
Such time mismatch is attributed to electronic delays on the laser interconnections. 
It was accounted in the theoretical model by means of an unbalance in the lasers parameters.
The corresponding measured pump currents obtained for the case of only one laser with feedback is given in Fig. \ref{fig:Fig11}.
From this data we extract the value between $30$ and $100\,\mu A$ for the spikes in the currents. 
These are fluctuations at least one order of magnitude bigger than any thermal or quantum noise current fluctuation.

%
The result for a pair of synchronized drop is given in Fig. \ref{fig:Fig12} and \ref{fig:Fig13}.  
Again there is the time mismatch of nearly 5 ns, which correspond to $1/10$ of the feedback time. 
They show clearly that the signals do not drop on exact time coincidence. 
However, on the coarse grained (many nanoseconds) time scale both laser powers drop always together. 
The comparison with the model here, is excellent when we substitute $J_0=20\,mA\,$, obtaining 
an excursion of current variation equivalent to the experimental value of $\pm 30\,\mu A\,$.
%

%

\section{The onset of Synchronism}

\begin{figure}[!hbtp]
	\centering
	\subfigure{
		\includegraphics[width=4.0cm,height=4.0cm]{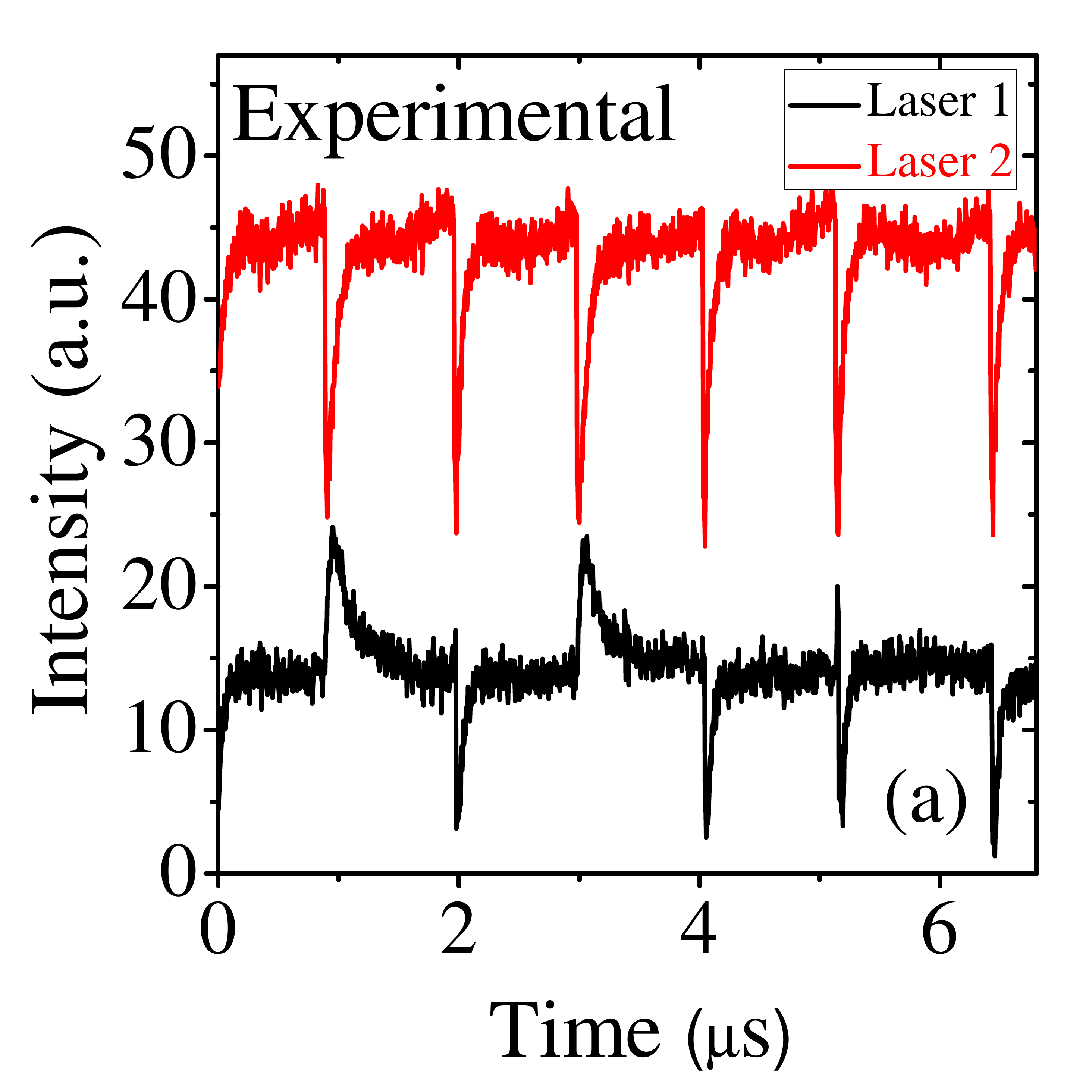} 
		\label{fig:Fig14a} } 
	\subfigure{
		\includegraphics[width=4.0cm,height=4.0cm]{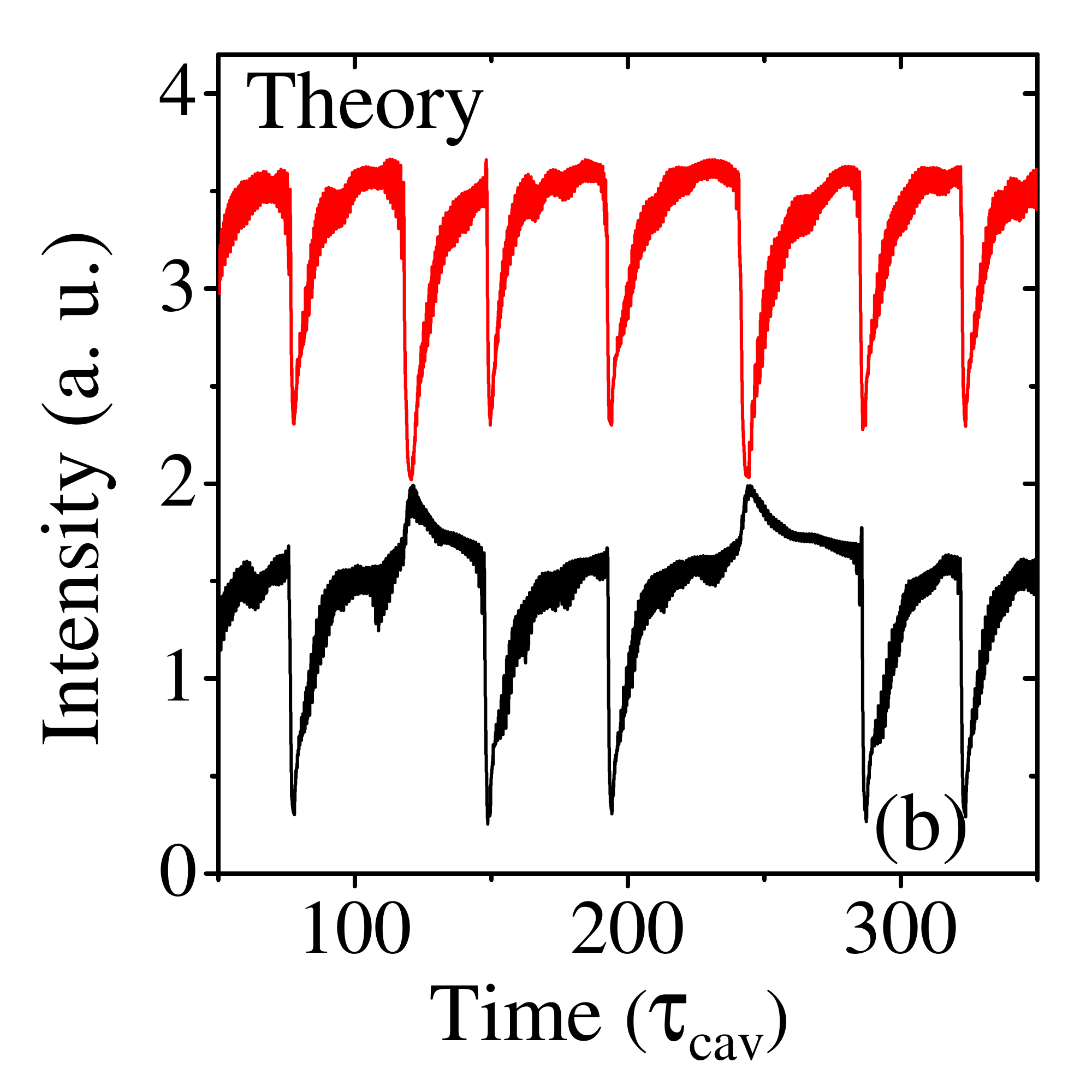}
		\label{fig:Fig14b}}
	\caption{Output power of the lasers coupled in parallel and having optical feedback. 
		The pump current and feedback strengths were set to partial synchronism of LFF. 
		Top line is Laser 2 and bottom Laser 1. (Power scales were displaced for visualization). 
		(a)Experimental and
		(b) Numerical integration from theory. 
		Whenever a coincidence drop is missed there is power jump up in one of the lasers.}
	\label{fig:Fig14}
\end{figure}

When the two lasers have optical feedback each one can manifest 
LFF power drops which, in general, are uncorrelated. 
Still the small anti-phase fluctuations are present. 
The onset of LFF synchronizations in time series is shown in Fig. \ref{fig:Fig14a}. 
The two lasers have LLF 
but only a partial number of LFF drops in synchronism is observed. In these cases,  instead of an increase of one 
laser power at the expenses of the big drop in the other one, both lasers drop power together.  
Figure \ref{fig:Fig14b} shows a numerical time series giving a 
segment calculated with the parameter $\eta$ varying in this intermediate synchronism regime. 

The rate equation model reproduces this dynamical 
condition as shown in Fig. \ref{fig:Fig14b}.
%
As we choose appropriate value for the total current and feedback alignments, more and more 
cases occur where both lasers have near simultaneous irregular LFF power drop until all power drops  
synchronize in phase with a small time mismatch of a fraction of the feedback time. Both, the experimental and the numerical theoretical curves 
showing intensity drops and current jumps during a synchronized LFF reveal a delay time between the two lasers. In the experimental system 
this delay was traced to the threshold of the lasers. The one with lower threshold, on the order of $2\% $, always dropped a few nanoseconds earlier.
Consistently, introducing a $1\%$ difference among the numerical values of threshold in the equations did reproduce the same type of delay. 
A quantitative inspection of LFF delay versus the other parameters was left for further studies.  Also, investigation of the role of significant electronic delays, by changing the length of the circuits wiring, will be left for a future work. In our current experiments, cables of less than $10$ cm restricted the possible delay to less than a nanosecond. 
Therefore, our observed delays, associated to threshold parameters, have physical origin on the carriers electronics in the junctions.

\section{On the phase of the chaotic intensity pulses and the optical phase of the fields}

The concept of synchronized chaotic optical oscillators described here deserves clear distinction from the concept of 
synchronization by frequency entrainment  in optical clocks. This last case demands that the two optical 
fields evolve in time with locked phase as 
\begin{eqnarray}
E_{j}(t) &=& |E_{0j}(t)|\exp{[- i\phi_{j}(t)]}\;, 
\label{eq:ephase}
\end{eqnarray}
with $\phi_{1}(t)-\phi_{2}(t) = constant\,$.

Most of chaos synchronization among lasers do not obey such condition. This is the case here.
Both in the experimental and numerical cases, we focus on the light intensity dynamics obtained 
from the squared field amplitude. 
The irregular pulses described by $I_{1}(t)$ and $I_{2}(t)$ 
are the result of a time averaging over optical periods and even more, averaging over detection time filtering.
Therefore, we are dealing with phases on intensity variations rather than amplitude oscillations. 
The phases that we get for the dynamical variables in their phase spaces can appear as 
locked when chaos synchronization is 
attributed to the coupled dynamics but there is no locking of the optical phase. 
Specifically, the averaged winding numbers calculated for the two slowly varying field envelopes
in our rate equations became equal in the condition we call synchronous.
Detailed discussions on the phase synchronization in coupled chaotic oscillators can be found in 
\cite{pikovosky1997,rosenblum1997}. 

\begin{figure}[!hbtp]
	\centering
	\subfigure{
		\includegraphics[width=4.0cm,height=3.8cm]{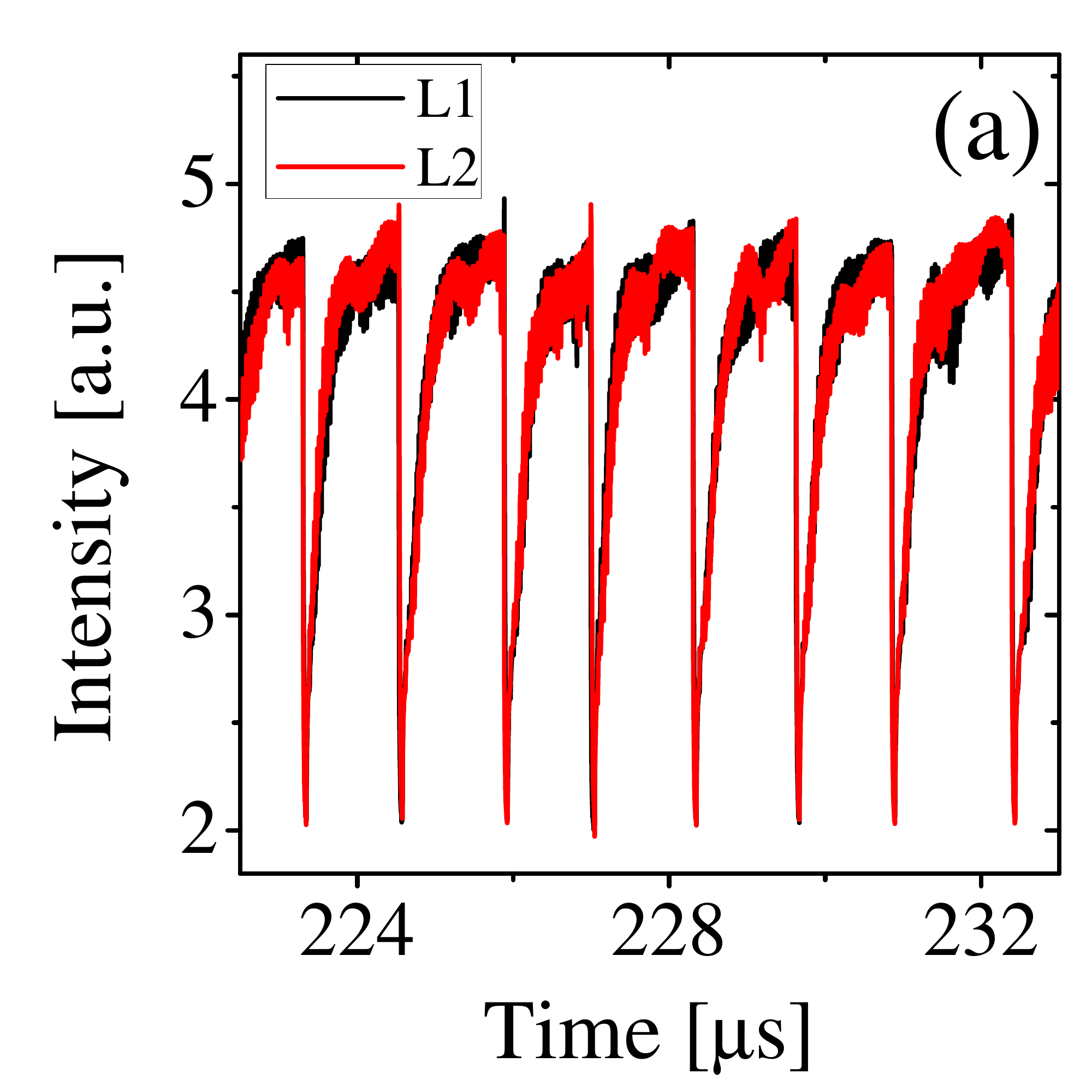} 
		\label{fig:Fig15a} } 
	\subfigure{
		\includegraphics[width=4.0cm,height=3.8cm]{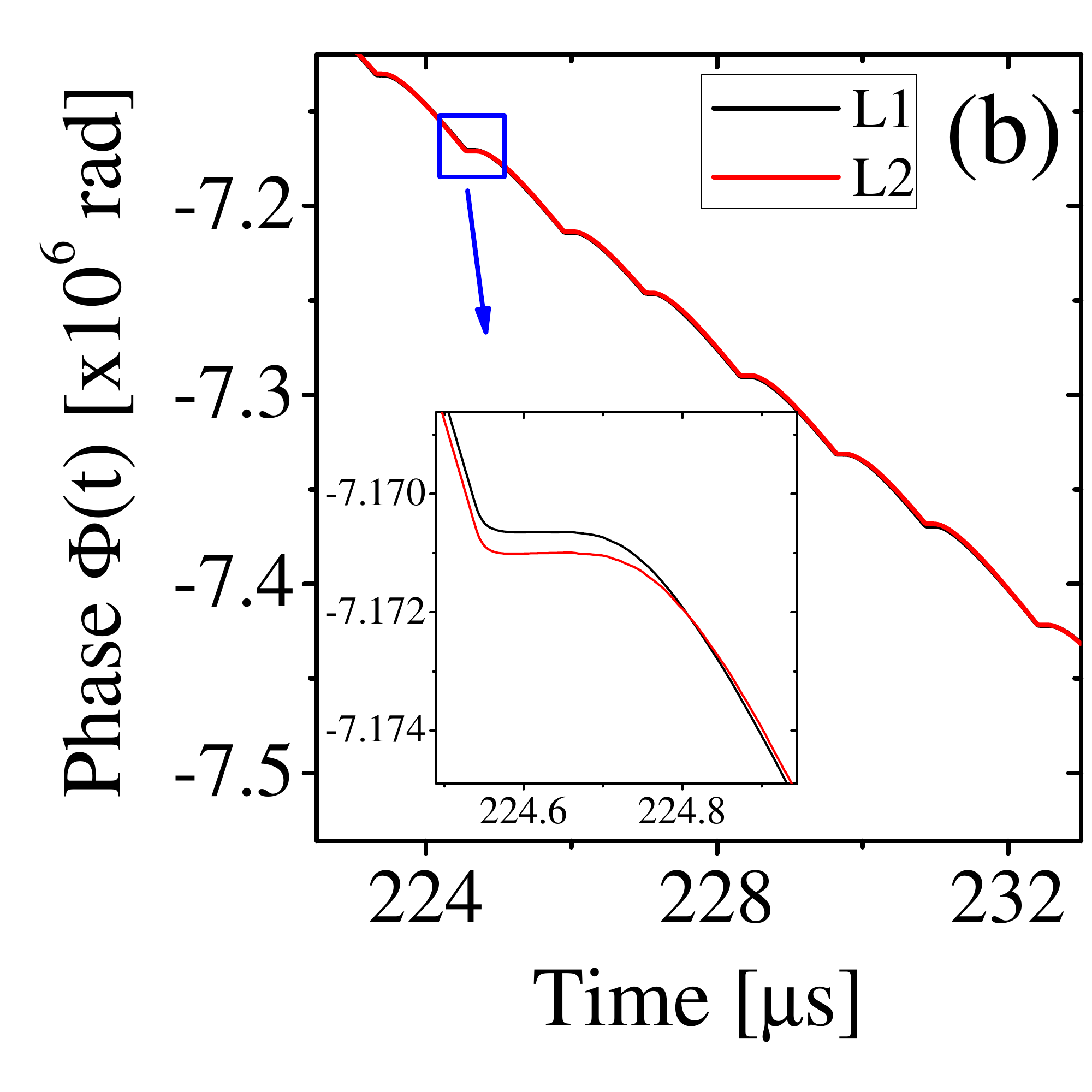}
		\label{fig:Fig15b}}
	\subfigure{
		\includegraphics[width=4.0cm,height=3.8cm]{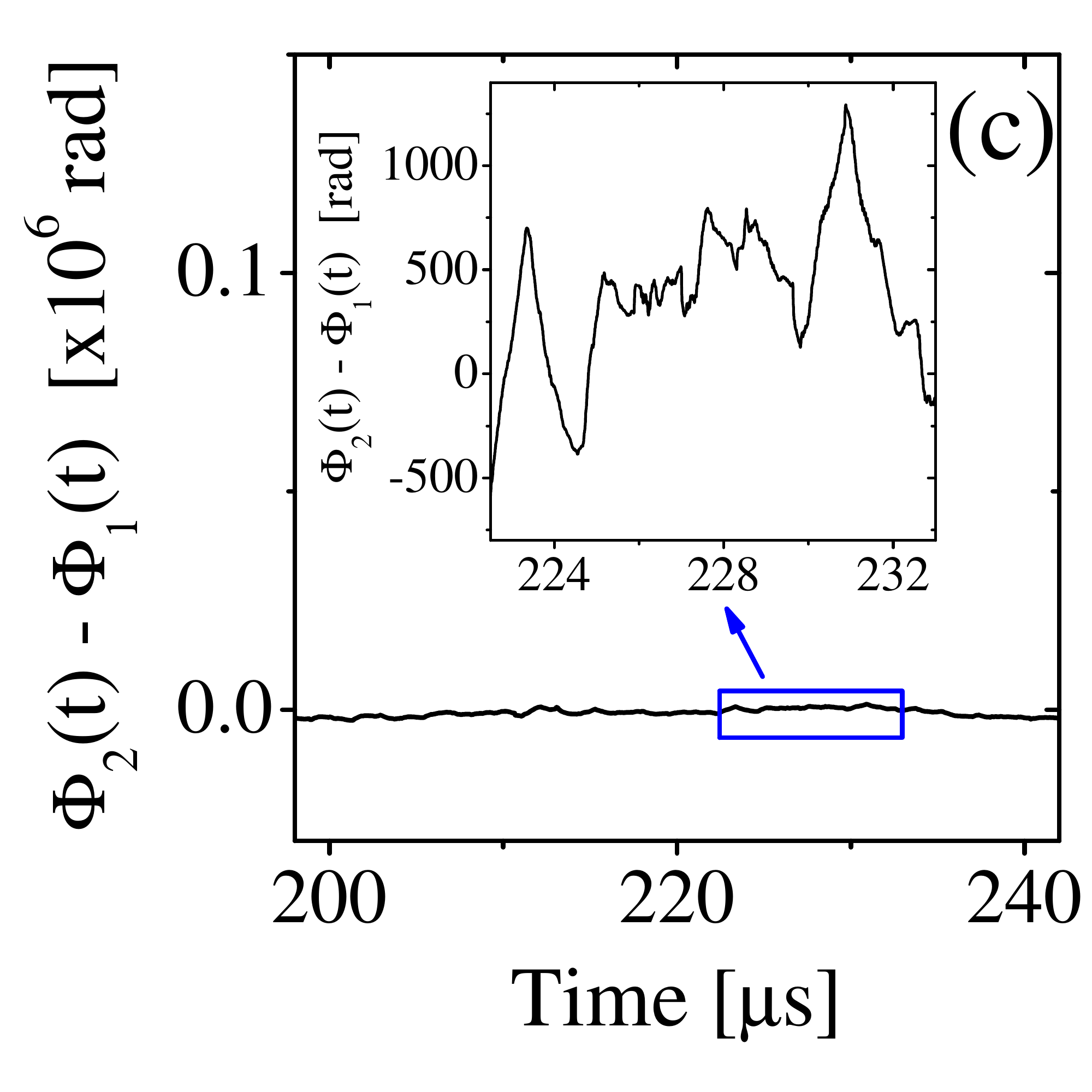}
		\label{fig:Fig15c}}	
\subfigure{
		\includegraphics[width=4.0cm,height=3.8cm]{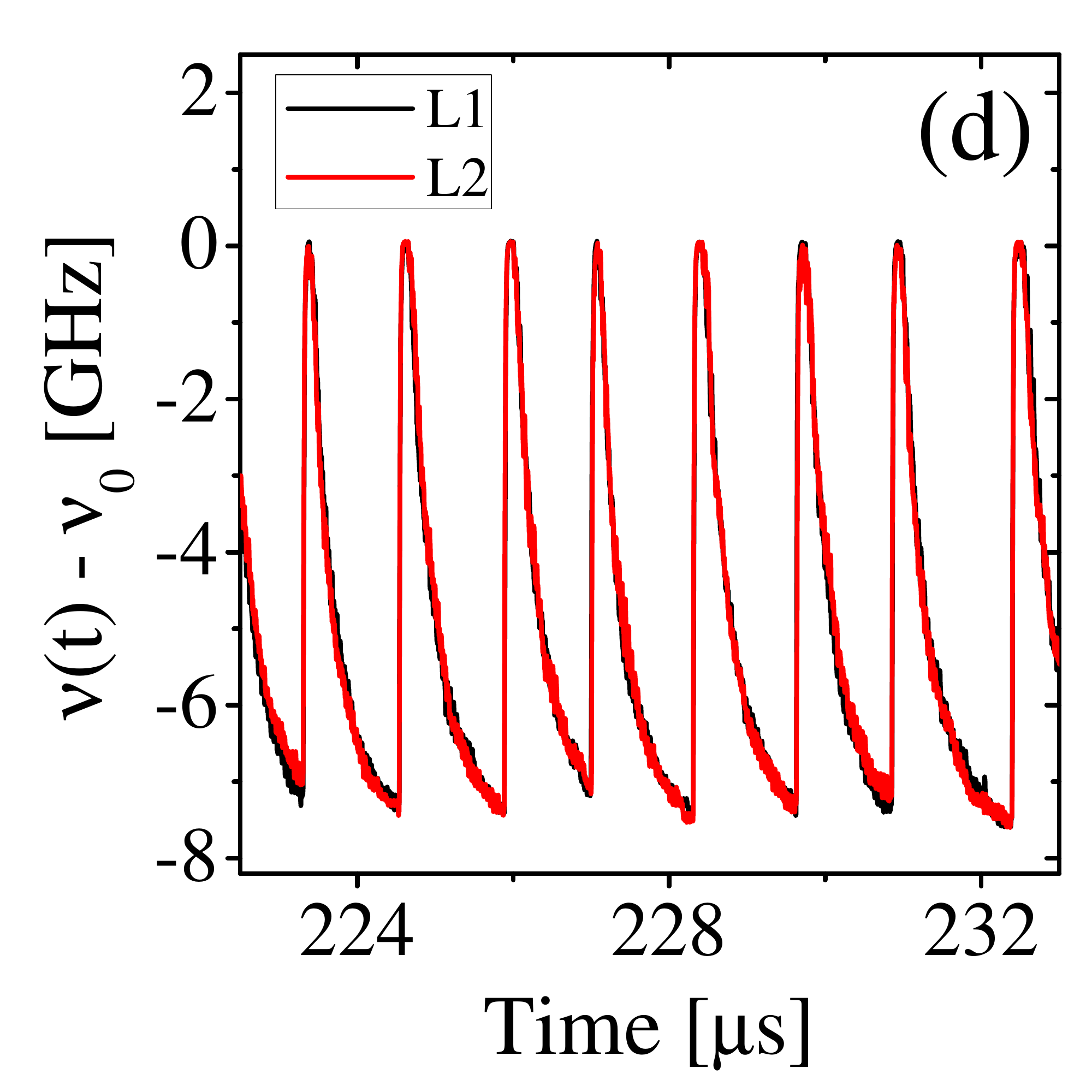}
		\label{fig:Fig15d}}
\subfigure{
		\includegraphics[width=4.0cm,height=3.8cm]{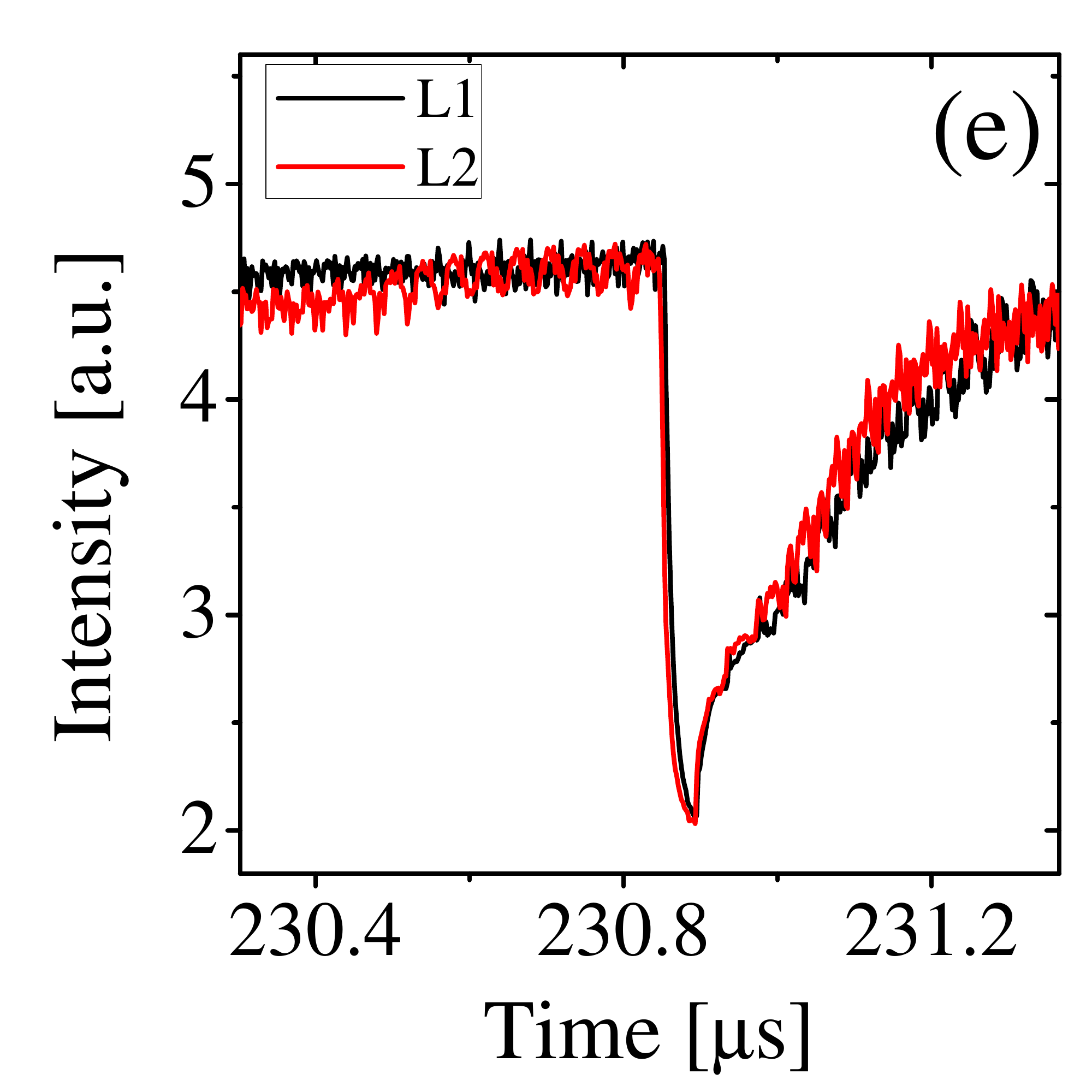}
		\label{fig:Fig15e}}	
\subfigure{
		\includegraphics[width=4.0cm,height=3.8cm]{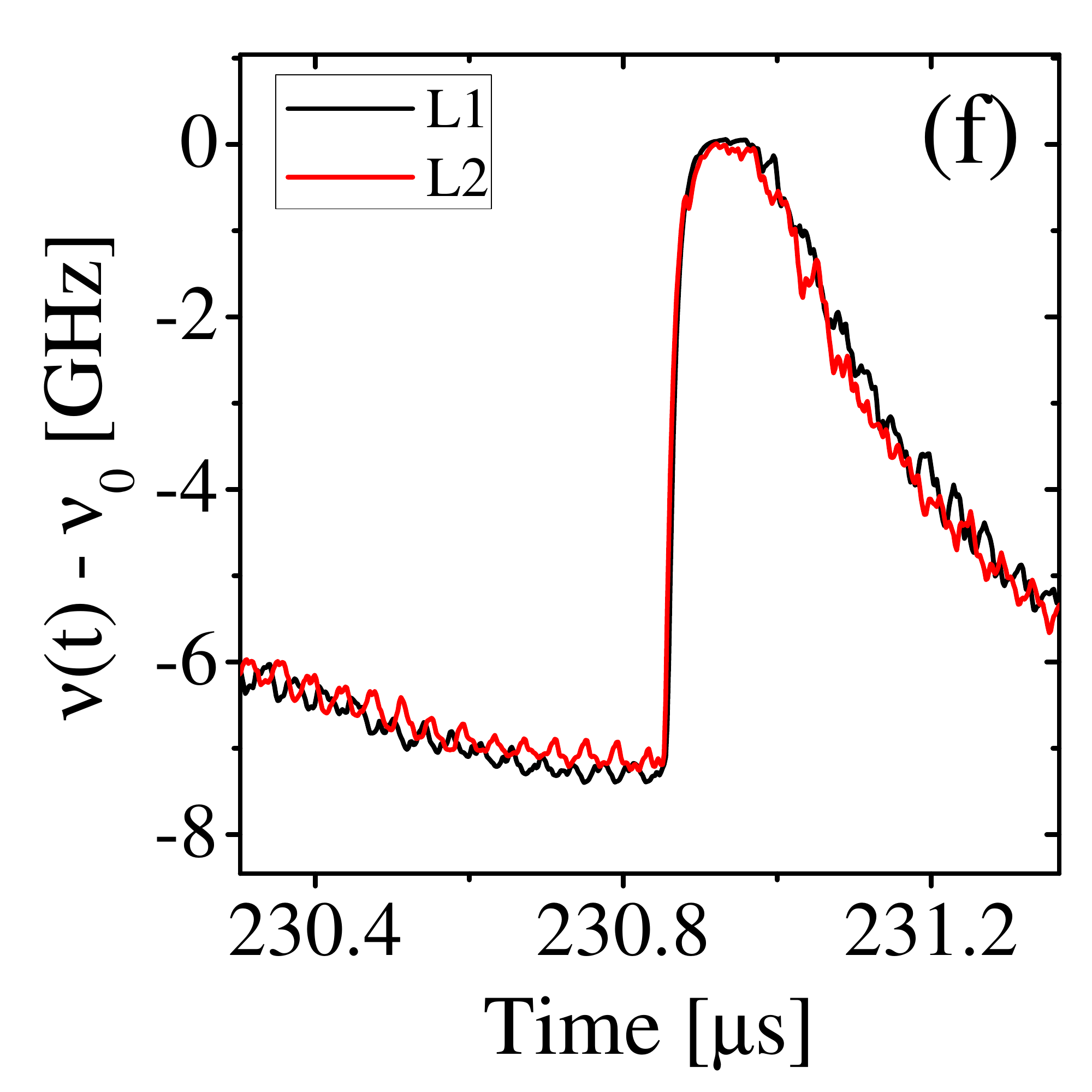}
		\label{fig:Fig15f}}	
		\caption{Numerically calculated optical phase for the two coupled lasers. 
		  (a) Segment of the laser intensity for the sake of comparison, (b) Plot of the two phases evolving in time.
		  (c) Phase difference showing the large time scale synchronism. (d) 
		 Optical frequency excursion of the two lasers. (e) Detail of the intensity fluctuations closed to a synchronized drop and (f)  the instantaneous frequency of the two lasers}
	\label{fig:Fig15}
\end{figure}

We can investigate the behavior of the optical phases from the numerical time series for the complex 
field amplitudes. 
Figures \ref{fig:Fig15a} to \ref{fig:Fig15f} show the optical phases and frequencies calculated from 
Eqs. (\ref{eq:LKeq1})-(\ref{eq:LKeq3}) when the lasers have synchronized LFF. 
It is important to notice that in Fig. \ref{fig:Fig15b}
the phases evolve with respect to the solitary laser phases, given by $\omega_{0}t],$, 
while most of the time the lasers have a red shifted frequency near the maximum gain condition. 
From the calculated slope we infer that circa 300 external cavity modes, separated by $\Delta\omega= 25 MHz$, participate in their itinerant dynamics \cite{sano}. 
The difference among the two phases is near zero in a rough coarse grained time scale as seen in Fig. \ref{fig:Fig15c}. 
The numerical time derivative of the phases give the instantaneous frequency. The value for each laser is  shown in Fig. \ref{fig:Fig15d} where they appear superimposed.

%
%
During each pair of drops, that appear to be
simultaneous when viewed in a large time scale (tens of nanoseconds), the two lasers can have 
different optical phases with jumping excursions of thousands of radians. 
In Fig. \ref{fig:Fig15c}, the phase mismatch is shown in two short time intervals around LFF drops. 
Notice that the short time (tens of nanosecond) oscillations of the phases are partially synchronized in anti phase.
The single drops, with a shorter time scale permits the observation of the anti correlated fast oscillations both in the intensities and in the 
optical phase at different time intervals before the drop. The ($7.5 GHz$) irregular red frequency chirping during each LFF 
drop and recovery cycle is clearly observed in Fig. \ref{fig:Fig15f}.
In all cases a precise optical synchronism is never attained, consistent with our experiments.  
%
\section{On the Correlation Functions}

\begin{figure}[!hbtp]
\centering
\subfigure{
\includegraphics[width=4.0cm,height=4.0cm]{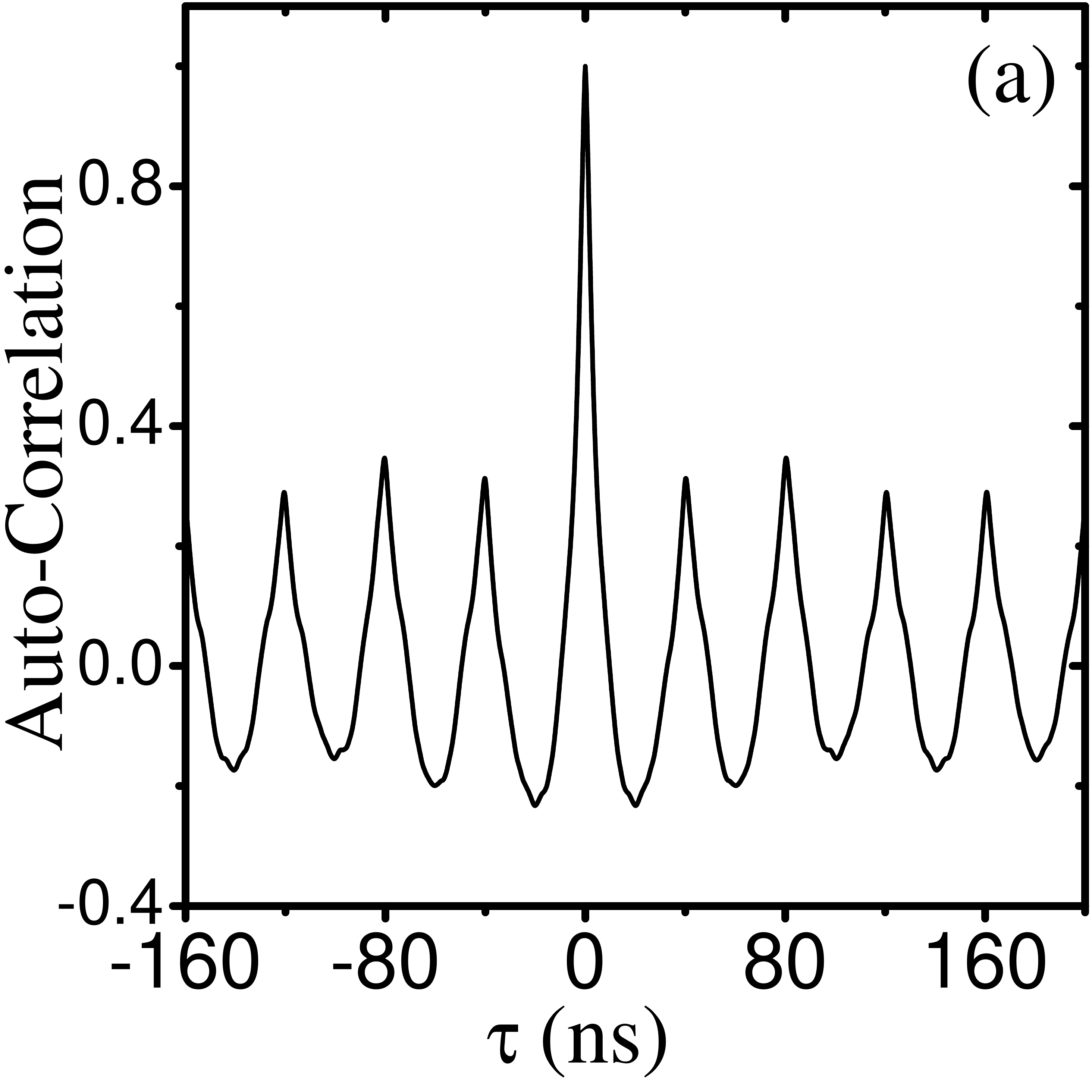} 
\label{fig:Fig16a} } 
\subfigure{
\includegraphics[width=4.0cm,height=4.0cm]{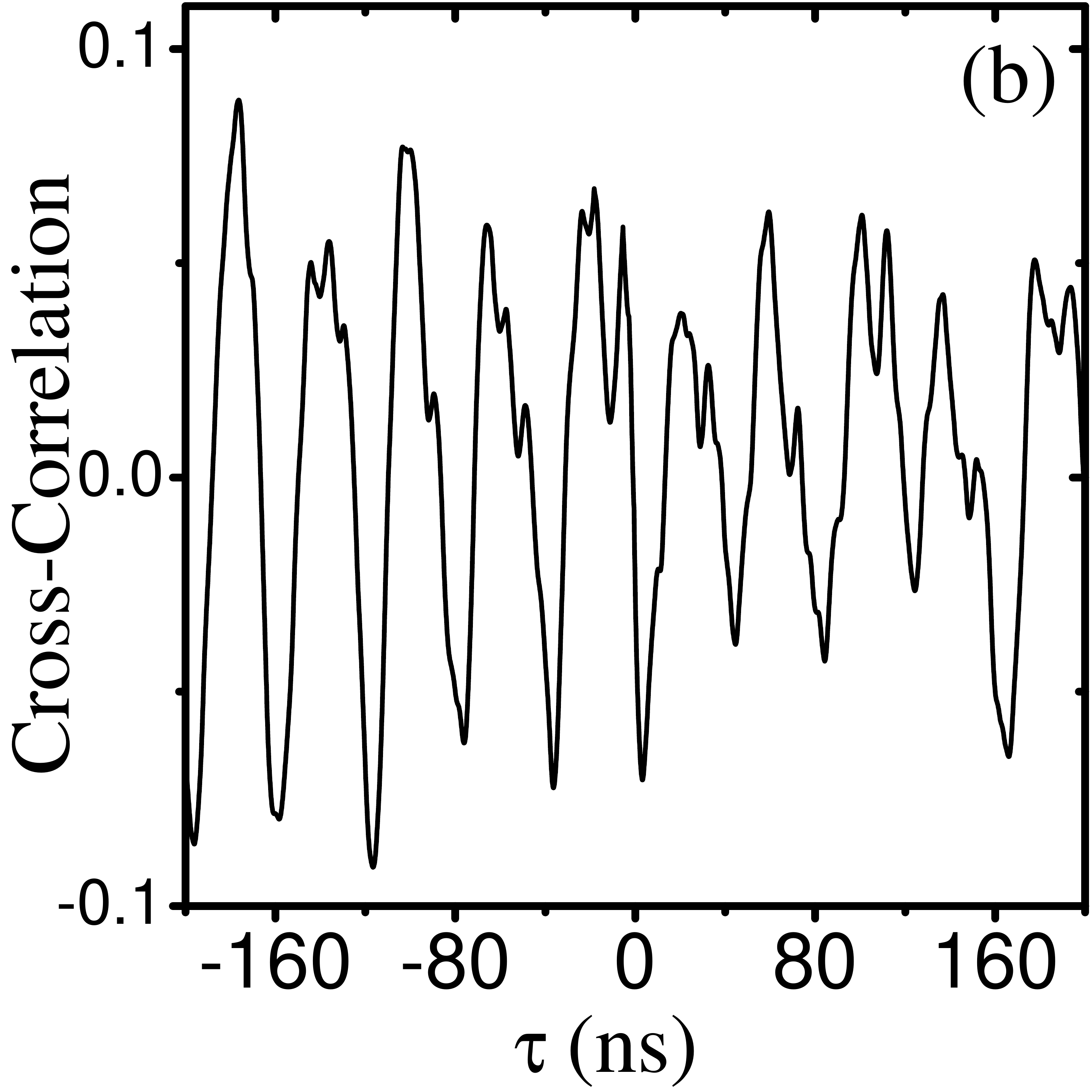}
\label{fig:Fig16b}}
\caption{Experimental correlations of the lasers coupled in parallel and having optical feedback. 
		(a) Autocorrelations of one laser. 
		(b) Cross correlation showing negative values due to anti phase fluctuation contributions, superimposed to 
		large time scale positive contribution due in phase fluctuations.}
\label{fig:Fig16}
\end{figure}

\begin{figure}[!hbtp]
\centering
\subfigure{
\includegraphics[width=8.0cm,height=4.0cm]{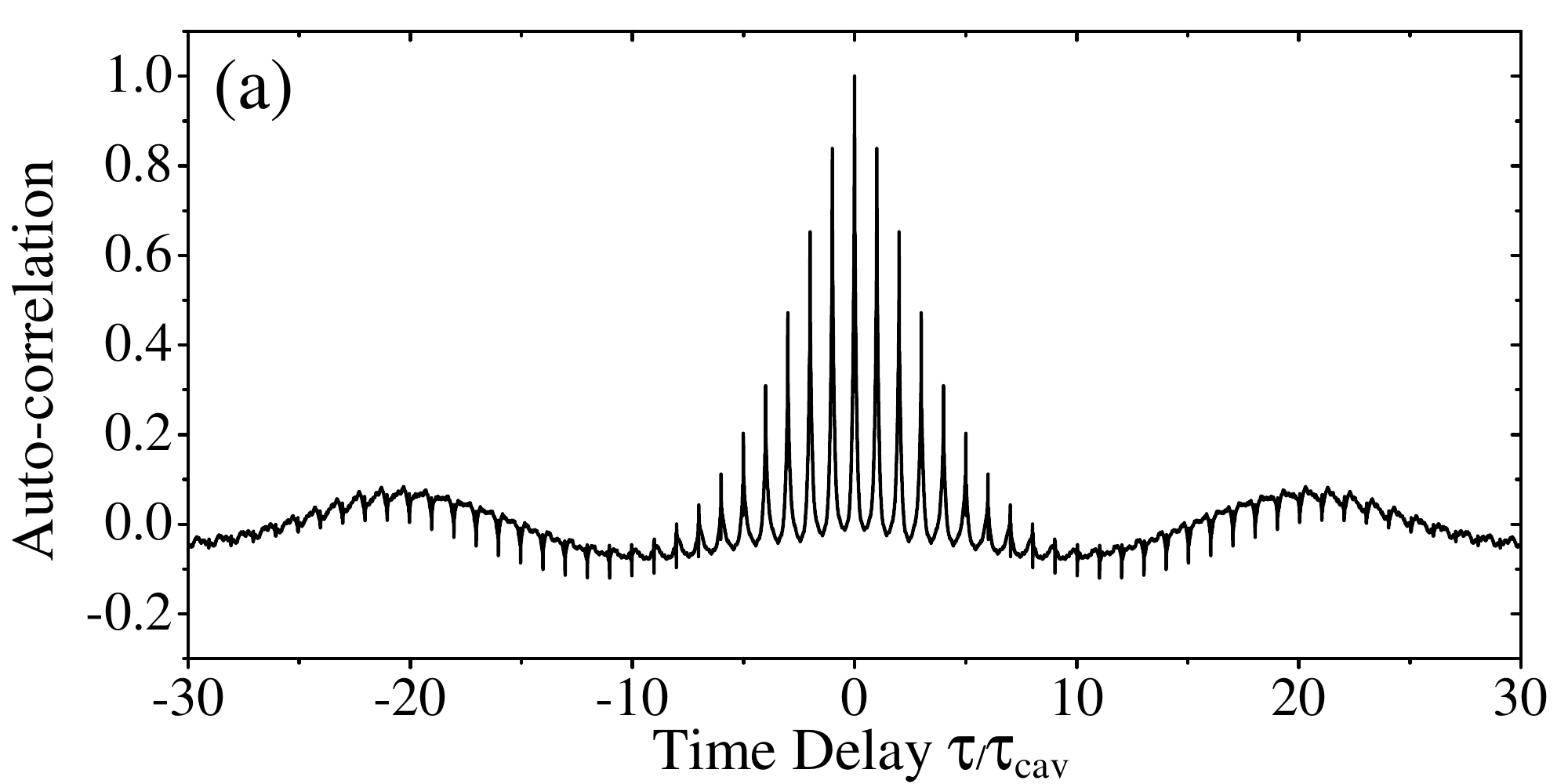}
\label{fig:Fig17a}}
\subfigure{
\includegraphics[width=8cm,height=4.0cm]{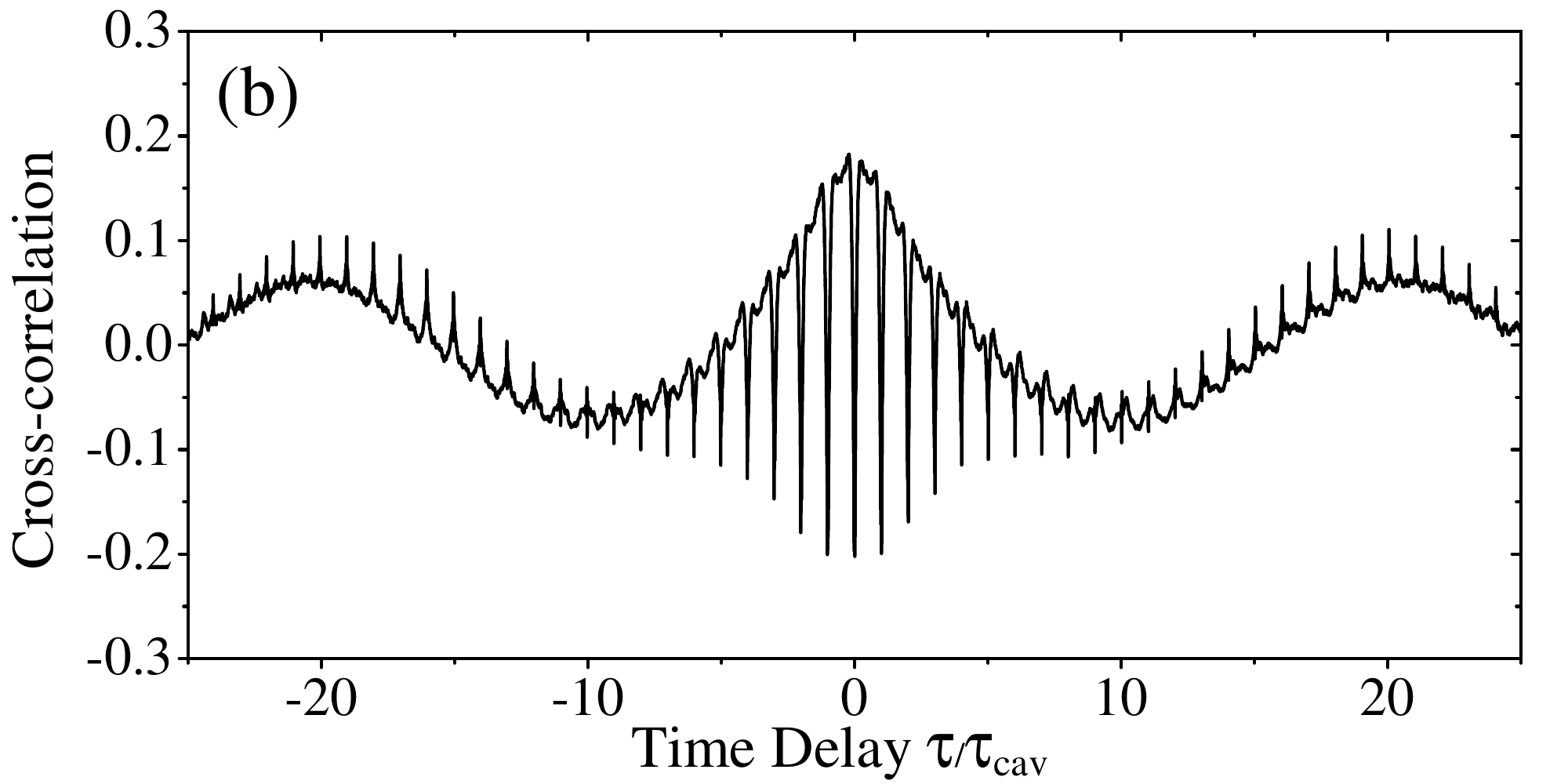}
\label{fig:Fig17b}}
\caption{Numerical correlation functions: (a) Auto correlation of laser 1. 
		(b) Cross correlation showing the down going anti phase fine structure superimposed on the in phase synchronized LFF.
		Filters were used to enhance  the proportions of anti phase (negative contribution)  
		and in fase (positive contribution) signal visible.}
\label{fig:Fig17}
\end{figure}

Proper setting for synchronization could give long data series containing more than $10^5$ LFF
events missing no pair of simultaneous drops.
The  competition for pump energy still exists, but only in short time scale intervals.
Fig. \ref{fig:Fig16a}, shows the autocorrelation function calculated from experimental series measured for one laser to compare 
with the experimental cross correlation between the two synchronized lasers in Fig. \ref{fig:Fig16b}. 
%
The theoretical autocorrelation function for one laser and cross correlation for the two synchronized 
lasers are shown in Fig. \ref{fig:Fig17}.

%
Autocorrelation and crosscorrelation functions of signals having more than one dominant 
(not necessarily perfectly periodic) period 
need interpretation when one wants to identify in-phase, positive correlated, 
and anti-phase, negative correlated signals.
To clarify how in-phase synchronized signal
events in slow time scale coexisting with 
some anti-phase oscillations of individual units in a coupled system manifest in cross correlation 
functions, we did numeric calculations using non chaotic time series extracted from the functions 
\begin{eqnarray}
X(t) &=& A\cos(\Omega_{1}t)+a\sin(\omega_{1}t)\;, 
\nonumber\\
Y(t) &=& B\cos(\Omega_{2}t) +b\sin(\omega_{2}t)\;.
\label{eq:XY}
\end{eqnarray}
When $\Omega_1/\Omega_2$ and $\omega_1/\omega_2$ are not rational,  the cross correlation between 
these signals is zero.
Interesting results appear when we consider $\Omega_1=\Omega_2=\Omega$ and $\omega_1=\omega_2=\omega$ 
and change the values of the coefficients. The cross correlations of these signals will give clues to what we 
get for the cross correlations from the nonlinear equation for the electrically coupled lasers having optical feedback.

\begin{figure}[!hbtp]
	\centering
	\subfigure{
		\includegraphics[width=6cm,height=4cm]{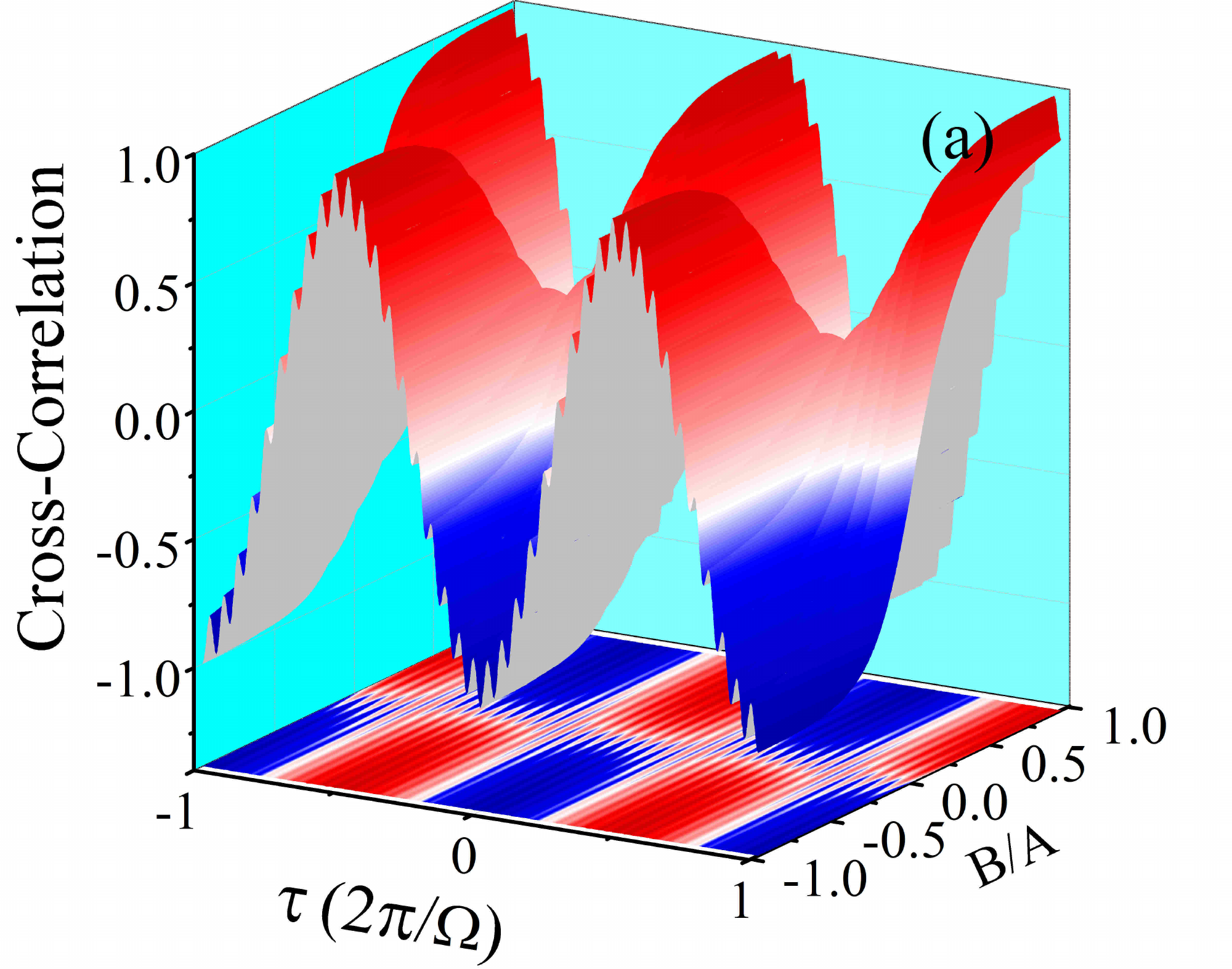} 
		\label{fig:Fig18a}} 
	\subfigure{
		\includegraphics[width=6cm,height=4cm]{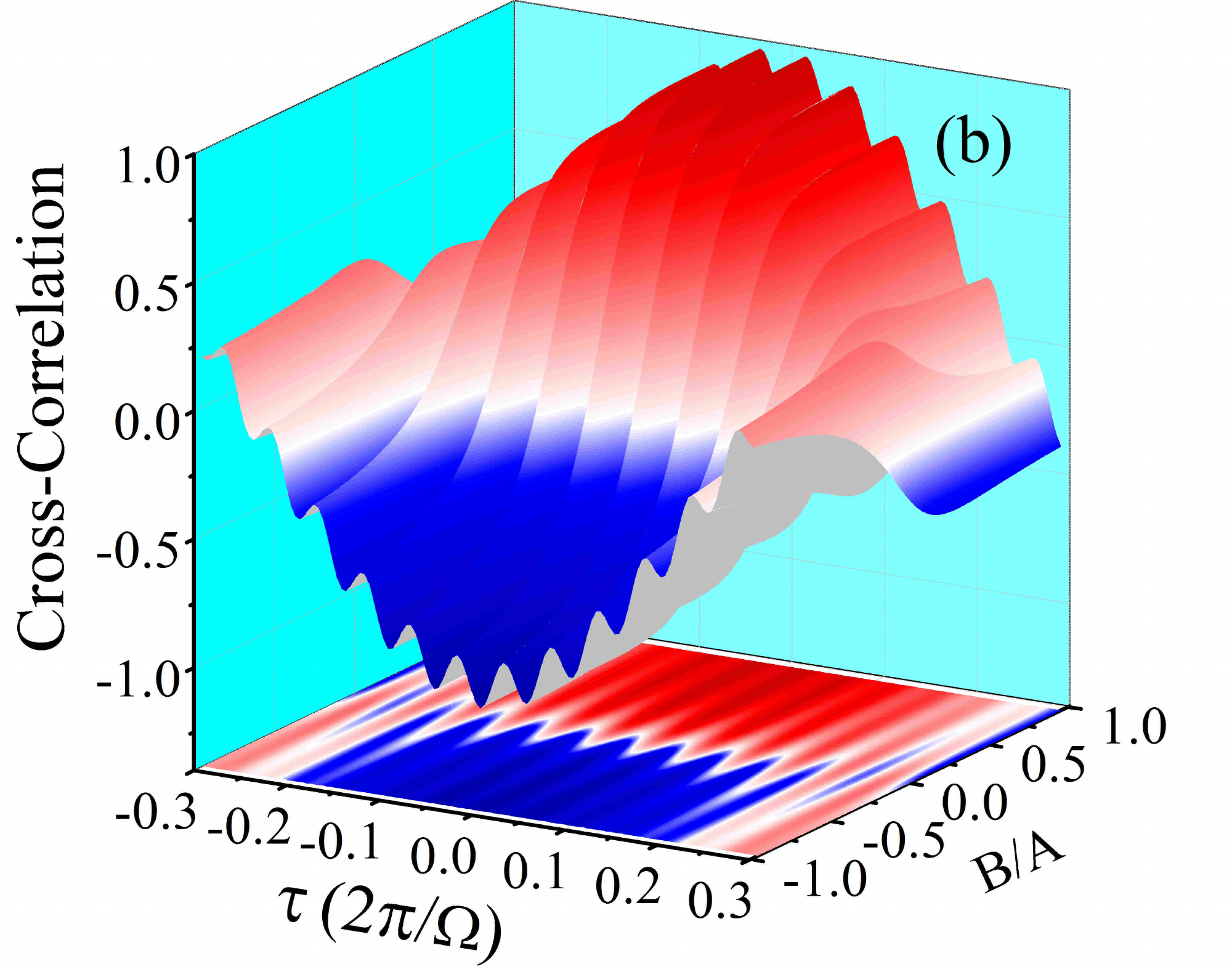} 
		\label{fig:Fig18b} } 
	\caption{Numerical cross correlation obtained when two simple signals are composed by harmonics in 
		phase and anti phase, with different frequencies and having variable amplitudes.  
		(a) 3D map showing the dominant anti phase term contribution evolving to the in 
		phase cross correlation when $B$ in Eq. (\ref{eq:XY}) varies from $-A$ to $A\,$. 
		(b) Detail of (a) showing the undulations associated to anti phase overriding the wide positive correlation curves.}
	\label{fig:Fig18}
\end{figure}

\begin{figure}[!hbtp] 
	\centering  
	\subfigure{
		\includegraphics[width=4.0cm,height=4.0cm]{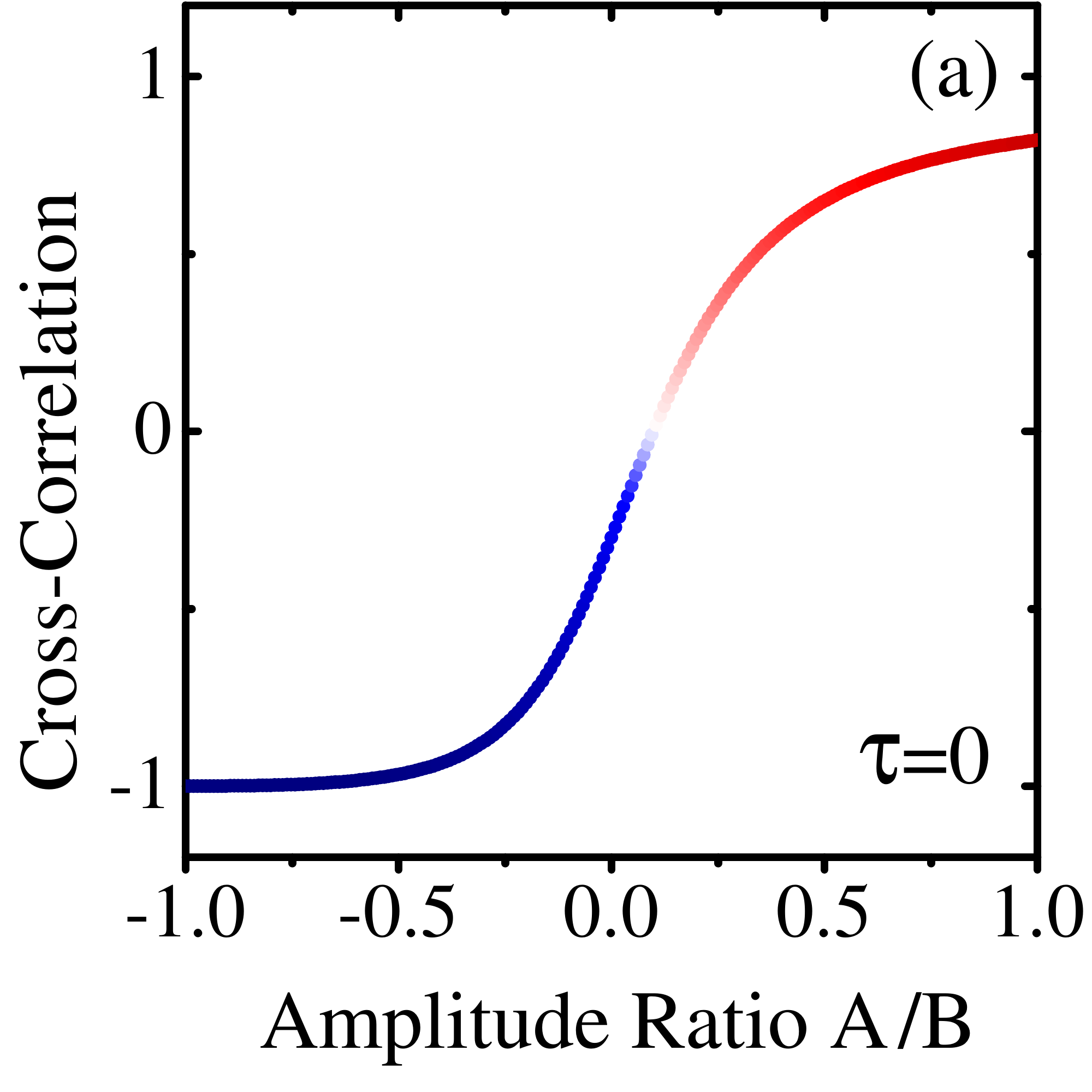}
		\label{fig:Fig19a}}
	\subfigure{
		\includegraphics[width=4.0cm,height=4.0cm]{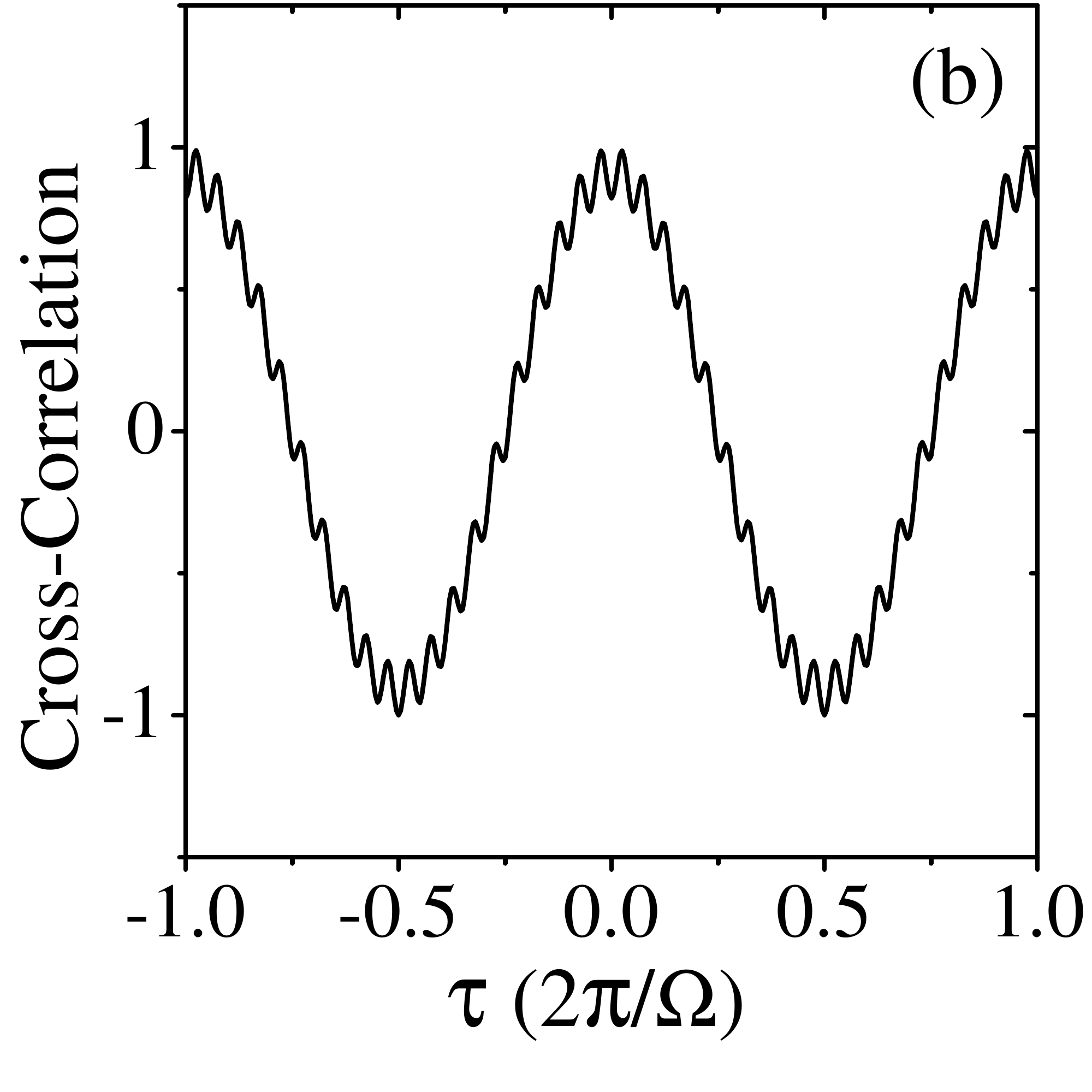}
		\label{fig:Fig19b}}
	\caption{Details of the numerical cross correlation obtained when two simple signals are composed by harmonics in 
		phase and anti phase with different frequencies and having variable amplitudes.  
		(a) Variation of the zero time crosscorrelation showing the excursion from fully antiphase, $-1$ to 
		almost complete in phase near $+1$. The value $+1$ is not reached because a small anti phase contribution is always present.
		(b) Correlation profile when $B=A\,$.}
	\label{fig:Fig19}
\end{figure}

Taking a total time much larger than $1/\Omega$ and $1/\omega\,$, 
the analitical expressions for the auto correlations $C_{1}$ and cross correlation $C_{12}$ at $\tau=0$ 
between the signals become
\begin{eqnarray}
C_{12}(\tau=0) &=& \frac{\langle\Delta X\Delta Y\rangle}{\sqrt{\langle({\Delta X})^2\rangle\langle{(\Delta Y})^2\rangle}} 
\nonumber\\
&=& \frac{(AB+ab)}{\sqrt{(A^2+a^2)(B^2+b^2)}}\;.
\end{eqnarray}
Our interest is when $\omega$ is at least one order of magnitude bigger than $\Omega$ and the two 
time scales have opposite correlations. This is the case when $AB>0$ and $ab<0\,$, for example.
Let us take $A$ fixed and positive, and increase the value of $B$ starting from zero. 
If $A\gg a$ and $B\gg b$ we have $C_{12} \rightarrow +1$ and, conversely, if $A\ll a$ and $B\ll b$ we have 
$C_{12} \rightarrow -1$.
To visualize numerically calculated figures, we took $A=105$, $a=-b=33$, $\Omega=2\pi$ and $\omega=20\times 2\pi$.
Varying $B$ from zero to the value of $A$ we get the  calculated cross correlations shown 
in Figs. \ref{fig:Fig18} and \ref{fig:Fig19}.
For small $B$ the anti-phase contribution from the two fast sinusoidal signals dominates. 
This is seen in the negative portion of the cross correlation at a zero time. 

The cross correlation of the laser dynamics in chaos have qualitatively the same shape seen in 
the respective figures above.  We can therefore interpret the onset of LFF synchronism obtained 
in our experiments by inspection of these correlation functions. 
%

%

\section{Related Dynamics in Multimode Lasers}

The dynamics of coupled monomode laser oscillators have general features common to the dynamics of single lasers with more than one field mode.  
Such common properties result form the fact that the number of dynamical variables in both systems can be the same. Two monomode lasers, 
like the ones described by our equations, before optical feedback is included, consist of a pair of three dimensional dynamical system: 
One complex field , $E_{i}(t)$ and one gain population $N_{i}(t)$ for each. A single laser like a VCSEL, when described as two polarization modes \cite{sanmiguel,giudicceVCSEL, sciamanna03},
also has two ortogonal fields, $E_{+}(t)$ and $E_{-}(t)\,$, competing for two populations $N_{+}(t)$ and $N_{-}(t)\,$. These are associated to the  different sub bands as proposed by San Miguel {\it et al.} \cite{sanmiguel}. The inter lasers coupling mechanism in our case is attributed to Kirchhoff's law as current conservation in the parallel circuit, 
while the two populations in the VCSEL model are coupled by spin flip mechanisms. 
The introduction of optical feedback in either system brings the infinite dimensional 
feature depending on time delay and the two subystems show LFF \cite{giudicceVCSEL}. Antiphase and inphase 
correlations among the two laser intensities $ I_{1}(t)$ and $I_{2}(t)\,$, as for $I_{+}$ and $I_{-}(t)$ in VCSELs, 
should therefore not be a surprise to be found in both systems. 
Also, intensity correlations between different spatial modes, both in edge emitting diode lasers and VCSELs, 
have been investigated long ago in Refs.\cite{khouryvcsel1,khouryvcsel2,khouryvcsel3,khouryvcsel4,khouryfourier}, 
where the spatial distribution of 
intensity noise was measured and intermode anti-correlations were responsible for total intensity noise reduction, 
sometimes below the shot noise level. 

The phase correlations in VCSEls are 
very well detailed in the paper by Sciamanna and coauthors \cite{sciamanna03}. These authors call attention to the need of more 
experimental work on the VCSEL dynamics and we hope to include such studies in our future research with coupled systems.
A detailed description of the 
optical feedback effects in Vertical-Cavity Surface-Emitting and Edge Emitting 
semiconductor lasers is given by Panajotov {\it et. al.} \cite{Vcsel2013}. Edge emitting are the types used and described 
in our experiments, but we have 
evidence for the same effects with Vertical-Cavity Surface Emitting ones. 
Discussions referring to coexistence between fast antiphase oscilations within multimode LFF 
dynamics can be found in the literature \cite{huyet99,sukov99}.

\section{Conclusion}

We have thus demonstrated in details how in-phase synchronized dynamical 
events in slow time scale can coexist with 
anti-phase oscillations between individual units in a pair of coupled lasers.
A remarkable observation, both in the experiments and in the simple theoretical model, 
is some enhancement of the anti-phase oscillations amplitude just before any sharply synchronized power drop. 
We also  
obtained numerical cases where the anti-phase enhancement before the drops appear on the frequency correlation.
The variation of correlation signals with time scale shown here for  
two lasers is relevant to the study of multi laser networks and their use for simulation of complex systems. 
Pairwise correlations within a complex network are known to be distinct from correlations among large groups of units \cite{Neurons2006,CorrelNeuronNet2011}, 
as well as different from the correlations in a single coupled pair system \cite{Edwards92}.
Detailed knowledge of the properties of pair units will enlarge the possibility to extract properties resulting from
the multi coupling topology in networks with many complex subsystems.

\acknowledgments

Work partially supported by Brazilian
Agencies: Conselho Nacional de Desenvolvimento Cient\'{\i}fico e Tecnol\'ogico (CNPq), 
Coordena\c c\~ao de Aperfei\c coamento de Pessoal de N\'{\i}vel Superior (CAPES), 
Funda\c{c}\~ao de Amparo \`a Pesquisa do Estado do Rio de Janeiro (FAPERJ) and
Funda\c{c}\~ao de Ci\^encia de Pernambuco (FACEPE).

\end{document}